\documentclass[sn-mathphys-num]{sn-jnl}

\usepackage{graphicx}%
\usepackage{multirow}%
\usepackage{amsmath,amssymb,amsfonts}%
\usepackage{amsthm}%
\usepackage{mathrsfs}%
\usepackage[title]{appendix}%
\usepackage{xcolor}%
\usepackage{textcomp}%
\usepackage{manyfoot}%
\usepackage{booktabs}%
\usepackage{algorithm}%
\usepackage{algorithmicx}%
\usepackage{algpseudocode}%
\usepackage{listings}%
\usepackage{comment}

\usepackage{CJK}
\usepackage{braket}
\usepackage{physics}
\usepackage[justification=justified]{caption}
\usepackage[numbers,sort&compress]{natbib}
\setcitestyle{super,open={},close={},comma}

\begin{document}

\title{`Interaction annealing' to determine effective quantized valence and orbital structure:\\ an illustration with ferro-orbital order in WTe$_2$}

\author[1,2]{Ruoshi Jiang}
\equalcont{These authors contributed equally to this work.}

\author[1,3]{Fangyuan Gu}
\equalcont{These authors contributed equally to this work.}

\author*[1,4,5]{Wei Ku}\email{weiku@sjtu.edu.cn}

\affil[1]{\orgdiv{School of Physics and Astronomy and Tsung-Dao Lee Institute}, \orgname{Shanghai Jiao Tong University}, \orgaddress{\city{Shanghai}, \postcode{200240}, \country{China}}}

\affil[2]{Department of Materials Science and Metallurgy, University of Cambridge, Cambridge CB3 0FS, United Kingdom}

\affil[3]{School of Materials Science and Engineering, \orgname{Shanghai Jiao Tong University}, \orgaddress{\city{Shanghai}, \postcode{200240}, \country{China}}}

\affil[4]{\orgdiv{Key Laboratory of Artificial Structures and Quantum Control} \orgname{Ministry of Education}, \orgaddress{\city{Shanghai}, \postcode{200240}, \country{China}}}

\affil[5]{\orgdiv{Shanghai Branch} \orgname{Hefei National Laboratory}, \orgaddress{\city{Shanghai}, \postcode{201315}, \country{China}}}

\abstract{Correlated materials are known to display qualitatively distinct emergent behaviors at low energy.
Conveniently, upon absorbing rapid quantum fluctuations, these rich low-energy behaviors can \textit{always} be effectively described by \textit{dressed} particles with fully quantized charge, spin, and orbital structure.
Such a powerful and simple description is, however, difficult to access through bare particles used in most many-body computations, especially when fluctuations are strong such as in $4d$ and $5d$ compounds.
To decipher the dominant quantized structure, we propose an easy-to-implement `interaction annealing' approach that utilizes suppressed charge fluctuation through enhancing ionic charging energy.
We establish its theoretical foundation using an exactly treated two-site Hubbard model as a generic example.
We then demonstrate its applications with more affordable density functional calculations to a representative $3d$ Mott insulator La${_2}$CuO${_4}$ and a highly fluctuating $5d$ semi-metal WTe${_2}$.
In the latter, it reveals an emergent local electronic structure that makes possible an unprecedented explanation of several experimental observations.
Finally, we demonstrate the effectiveness of this approach in studying competing local electronic structures in functional materials.}

\maketitle

\section{Introduction}\label{intro}

Modern functional materials~\cite{Robert2001,Sahu2019} display a rich variety of qualitatively distinct behaviors~\cite{Dagotto2005}, sometimes even switchable under weak change of external conditions, such as temperature, pressure, or external fields~\cite{Kawai2018, Kushnireko2017, Hualei0516, Cao_TBG_2018}.
Generically speaking, such rich physical properties are almost always associated with the underlying many-body correlation in the electronic charge, orbital, spin, or lattice degrees of freedom~\cite{Anderson1972}.
Particularly, the ability to qualitatively switch the properties of these functional materials indicates the existence of multiple competing electronic structures with significantly different coupling strengths to various external conditions~\cite{Taillefer2010, Dagotto2005, Dagotto1994}.

For complex systems like these, fortunately, there generally exist effective descriptions capable of \textit{exactly} capturing the essential low-energy dynamics using only the most relevant \textit{quantized} degrees of freedom~\cite{Anderson1972,Schrieffer1966,Zaanen1988,Jiang2024}.
Such simplification originates from the discrete energies of states spanning the significantly reduced \textit{low-energy} subspace~\cite{Anderson1972}.
Consequently, the physically more relevant slow dynamics of the system is \textit{rigorously} mappable to those of a few dominant many-body objects with quantized valence and orbital structure, upon absorbing rapid quantum fluctuations into their internal structure~\cite{Schrieffer1966,Zaanen1988,Jiang2024}.
In explicit many-body treatments, such ``dressing'' is many-body in nature~\cite{Schrieffer1966,Zaanen1988,Jiang2024}, while in typical density functional theory (DFT)~\cite{DFT1,DFT2} calculations, these fluctuations are absorbed in ``tails'' of Wannier orbitals~\cite{Wei2002, Kanungo2014}.
The existence of such quantized effective description, for example, explains the great success of integer valence count in determining the stability of chemical compounds~\cite{electrostatic,chargetrans,PIE1,PIE2,Jiang_int_valence}.

Since these fully quantized \textit{dressed} objects are the building blocks of the low-energy many-body states, they are essential for an intuitive understanding of the low-energy electronic structure of strongly correlated materials.
On the one hand, their dynamics gives the observed elementary charge-, orbital-, and spin-excitations of the system.
On the other, their correlation and ordering define those of the system.
As such, their spatial structures correspond to the ``form factors'' of inelastic and elastic scattering measurements~\cite{Larson2007Nonresonant, Walters2009Effect, Peter2008Dynamical}.
Therefore, seeking these fully quantized dressed objects, as often exercised in chemistry, is not simply for convenience, but rather an important task to gain direct access to the essential low-energy physics.

However, such simple quantized effective descriptions are not easily accessible in standard many-body treatments, when the quantum fluctuation (or hybridization) is strong.
Indeed, in terms of \textit{bare} particles employed in most numerical many-body treatments, the results always contain non-quantized expectation values of charge, orbital, and spin structures~\cite{supplementary, Nelson2015}.
It is therefore highly desirable to seek a simple mean for direct access to such fully quantized dressed objects within the \textit{existing} computational frameworks.
Specifically, we aim to find such dressed objects that emerge below the ionic charging energy in functional materials.

Here, to address this important issue, we propose an `interaction annealing' approach easily adaptable in standard many-body computations.
We first establish its rigorous theoretical foundation using an exactly treated two-site Hubbard model as a generic example.
We then demonstrate its applications with more affordable DFT calculations via two representative examples, $3d$ Mott insulator La${_2}$CuO${_4}$ and $5d$ semi-metal WTe${_2}$ that hosts strong charge fluctuation.
For the latter, this approach reveals a dominant ferro-orbital order in the DFT ground state, with W ions in a $d^2$ spin-0 orbital-polarized configuration, which explains several experimental observations~\cite{Tao_PhysRevB2020,Zhou_AIPadv2016,shannon1976revised,Yang2021}.
Finally, we demonstrate the effectiveness of the proposed approach in studying competing local electronic structures, which functionalize many modern materials.

\section{Results}\label{results}

Figure~\ref{schematics} illustrates the main idea of our proposed `interaction annealing' approach.
Given a system with a dominant interaction that dictates the highest-energy dynamics of the system, for example intra-atomic Coulomb repulsion $U$, we aim at obtaining the well-quantized dressed objects that constitute its physically most relevant dynamics {\textit{slower than time scale} $2\pi/U$~\cite{Sakurai1994}.
To this end, one can suppress the fluctuations through adiabatically~\cite{Kato1950, Eshuis2012Electron, Pernal2018Electron} (slowly) increasing the interaction, until the bare particles of the fictitious system display well-defined quantization.
Owing to the adiabatic connection between low-energy emergent objects belonging to the same stable ``fixed point'' of slower dynamics~\cite{granas2003fixed} (in the sense of renormalization group~\cite{Nigel1992}),
the structure of the well-quantized dressed objects, $\tilde{c}_{i\mu}^\dagger$, in the real system can be revealed by the bare objects, $d_{i\mu}^\dagger$, of the fictitious system.
Below, we establish the theoretical foundation via an exact many-body treatment of a simple model, followed by demonstrations via two representative examples.

\subsection{Quantized effective description}

As a simple yet generic illustration, consider a two-site Hubbard model~\cite{Hubbard1963Electron, Arovas2022},
\begin{align}
H=-t\sum_{\mu} (c^{\dagger}_{1\mu}c_{2\mu}+ c^{\dagger}_{2\mu}c_{1\mu}) + U (c^{\dagger}_{1\uparrow}c^{\dagger}_{1\downarrow}c_{1\downarrow}c_{1\uparrow}+ c^{\dagger}_{2\uparrow}c^{\dagger}_{2\downarrow}c_{2\downarrow}c_{2\uparrow}), \label{HubboardH}
\end{align}
containing two electrons, where $t$ denotes the hopping of particles $c_{1\mu}^\dagger$ at site 1, of spin $\mu$ ($\uparrow$ or $\downarrow$), to site 2 and back, and $U$ the charging energy when a site is doubly occupied.
Let's examine the emergent low-energy effective theories~\cite{supplementary} of this system in its correlated ($4t \le U$) regime.

In this regime, states with doubly occupied orbitals are of high energy.
Upon decoupling these states from the remaining low-energy sector via a unitary transformation of the basis~\cite{chao1977kinetic,White2002,QuantumAlgebra, Sanchez-Barquilla2020} (denoting $-\mu$ as ${\bar{\mu}}$ and the site next to $i$ as $i^\prime$), 
\begin{align}
\tilde{c}^\dag_{i\mu}&\equiv\mathcal{U}^\dag c^\dag_{i\mu} \mathcal{U} \label{dressed_c}
\\
&= c^\dag_{i\mu} + (\sin{\frac{\theta}{2}})\frac{1}{2} [c^\dag_{i\mu} (c^\dag_{i\bar{\mu}} c_{i'\bar{\mu}}-c^\dag_{i'\bar{\mu}} c_{i\bar{\mu}}) +c^\dag_{i'\mu} (n_{i'\bar{\mu}} -n_{i\bar{\mu}})] \nonumber\\
&+(\cos{\frac{\theta}{2}}-1)\frac{1}{2}[c^\dag_{i'\mu} (c^\dag_{i\bar{\mu}} c_{i'\bar{\mu}}+c^\dag_{i'\bar{\mu}} c_{i\bar{\mu}})+c^\dag_{i\mu} (n_{i\bar{\mu}} +n_{i'\bar{\mu}})], \nonumber
\end{align}
into the dressed particles in a `many-body Wannier orbital', $\tilde{c}^\dag_{i\mu}$, with $\mathcal{U}=\exp(\frac{\theta}{2}~\sum_\mu\frac{1}{2}(c^\dag_{1\mu}c_{2\mu}-c^\dag_{2\mu}c_{1\mu})(n_{2\bar{\mu}}-n_{1\bar{\mu}}))$
with $\theta=\arctan{(\frac{4t}{U})}$, the resulting effective theory for the remaining low-energy subspace,
\begin{align}
\tilde{H}^{(t\ll U)}=\tilde{J}_{12} (\tilde{\mathbf{S}}_1\cdot\tilde{\mathbf{S}}_2-\frac{1}{4}), 
\label{H_largeU}
\end{align}
contains one dressed particle in each site $i=1,2$ interacting via anti-ferromagnetic super-exchange coupling of strength $\tilde{J}_{12} = \frac{1}{2}(\sqrt{U^2+(4t)^2} - U)$,
where $\tilde{\mathbf{S}}_i=\sum_{\mu\nu} \tilde{c}^\dag_{i\mu} \mathbf{\sigma}_{\mu\nu} \tilde{c}_{i\nu}$ denotes the total spin of the \textit{dressed} particles via Pauli matrices $\mathbf{\sigma}_{\mu\nu}$.
(See Methods~\ref{tllU} for the transformed full $H$ containing both of the decoupled low- and high-energy sectors.)

Notice that the low-energy dynamics of \textit{all} systems of different $U$ are \textit{exactly} described by the \textit{same} effective description, Eq.~\ref{H_largeU}, containing only \textit{fully quantized} charge, orbital, and spin degrees of freedom (just with different renormalized parameter $\tilde{J}_{12}$).
For example, consider a `realistic' system with intermediate $4t/U=1$.
For \textit{all} low-energy eigenstates \textit{below the scale of} $U$, the spin $\langle\tilde{S}^2\rangle=\frac{3}{4}$ of the dressed particles corresponds to a quantized spin-1/2~\cite{Sakurai1994}.
In essence, the simplification through quantized effective descriptions rigorously results from \textit{fully absorbing} the rapid fluctuations into the internal structure of the dressed particles $\tilde{c}_{i\mu}^\dagger$, such that dynamics of the latter is no longer sensitive to such fluctuation in their representative time scale (longer than $2\pi/U$).
The existence of such fully quantized effective description(s) is therefore a generic characteristic of quantum many-body systems and generally applies to materials with open $d$- and $f$-shells having significant intra-atomic interactions.

However, such a fully quantized picture is not directly accessible from properties of bare particles, as typically computed in most many-body calculations.
Indeed, for $4t/U=1$, the spin $\langle S^2\rangle\sim 0.64 < \frac{3}{4}$ of the bare particles in the ground state deviates significantly from a quantized value, due to strong charge fluctuation, $\langle c_{i\mu}^\dagger c_{i\bar{\mu}}^\dagger c_{i\bar{\mu}} c_{i^\prime\mu}\rangle\sim0.18>0$, involving bare particles oscillating between sites and temporarily double-occupying one site.

\subsection{Adiabatic connection around a dominant structure}

To find a simple approach to access such quantized effective description, one can take advantage of the ``adiabatic connection''~\cite{Kato1950, Eshuis2012Electron, Pernal2018Electron} between the low-energy subspace of systems that share identical structure of slow dynamics (c.f. Fig.~\ref{schematics}).
For example, the above `realistic' system ($4t/U=1$) and systems with larger $U$ all share the same Hamiltonian Eq.~\ref{H_largeU}, indicating \textit{qualitatively identical} low-energy subspace (composed of $\tilde{c}^\dag_{i\mu}$) between these systems, only with different degree of many-body dressing according to Eq.~\ref{dressed_c}.
In technical terms, this means that the renormalization of slow dynamics in these systems shares the same stable ``fixed point''~\cite{granas2003fixed} and therefore they are adiabatically~\cite{Kato1950, Eshuis2012Electron, Pernal2018Electron} connected.

Furthermore, in systems with larger $U$ (closer to the fixed point), charge fluctuations are systematically suppressed.
Therefore, in fictitious `interaction annealed' systems with large enough $U$, the dressed particles, $\tilde{d}_{i\mu}^\dagger$, would become well resembled by the corresponding bare particles $d_{i\mu}^\dagger$.

The above generic properties of many-body emergence together provide direct access to the symmetries and quantum numbers of the dressed particles $\tilde{c}_{i\mu}^\dagger$ of a system, through `adiabatically annealing' (slowly increasing) the strength of the dominant interaction (e.g. $U$) until a clear quantized structure is obtained from the bare $d_{i\mu}^\dagger$.
Indeed, in the above representative example, as $U$ grows from $4t$ to $20t$ the charge fluctuation, $\langle c_{i\mu}^\dagger c_{i\bar{\mu}}^\dagger c_{i\bar{\mu}} c_{i^\prime\mu}\rangle$ reduces from 0.18 to 0.05, displaying the expected $t/U$ decay.
Consistently, $\langle S^2\rangle$ grows from 0.64 to 0.74, approaching the value $\frac{3}{4}$ of a well-quantized spin-1/2, exactly revealing the desired \textit{fully quantized} structure of the dressed particles in the `realistic' system.

\subsection{Practical applications under uncertainty of the dominant interaction}

Recall that the proposed approach aims at obtaining the emergent objects resulting from a ``dominant interaction'' of the system.
Without loss of generality, the above model example has demonstrated the conceptual \textit{existence} of such effective description via fully \textit{quantized} emergent objects, and furthermore the robust \textit{adiabatic connection} between systems of different strength of the dominant interaction, \textit{as long as} they belong to the same fixed point (namely the dominance of the interaction remains).

In practice, however, which interaction dominates the highest-energy dynamics of a system might not be obvious.
(As an example, see Methods~\ref{tggU} for the other fixed point of the model system.)
In that case, instead of a super-expensive full-blown many-body RG flow~\cite{Jiang2024, Jiang2025Emergence}, the proposed approach offers a much affordable route for practical studies.
One can simply explore various potential dominant interactions via the proposed approach and compares the corresponding emergent objects against experimental observations.
(See Section~\ref{ATCES} for a demonstration.)
Given the qualitative distinct characteristics of the emergent objects from different dominant interactions, it is unlikely that more than one of such emergent objects can be consistent with all the observed physical properties of the systems of interest.

\subsection{Application with density functional theory: 3$d$ Mott insulator La$_2$CuO$_4$}\label{AWDFT}

The above `interaction annealing' approach is supported by \textit{generic} emergent properties of many-body systems and is thus generally applicable to all many-body calculations using bare particle representation, such as variational Monte Carlo~\cite{yokoyama1987variational}, density matrix renormalization group~\cite{schollwock2005density}, and tensor network~\cite{evenbly2011tensor}.
We now demonstrate with two examples its applicability to more affordable DFT~\cite{DFT1,DFT2} calculations.

Considering the representative Mott insulator La$_2$CuO$_4$, we employ the standard LDA+$U$ implementation~\cite{Anisimov1993} that employs an effective Hartree-Fock treatment~\cite{Anisimov1997}.
This self-interaction corrected implementation~\cite{Anisimov1993} accounts for the effective self-interaction in the double counting term, such that the energetic sequence relative to the ligand orbitals is retained upon suppressing charge fluctuations via enlarged $U$.

With the realistic $U=8$ eV~\cite{Anisimov2004}, the one-body density matrix of the anti-ferromagnetic ground state gives a fluctuating $\langle d^\dagger_\downarrow d_\downarrow\rangle\sim 0.31$ for Cu $d_{x^2-y^2}$-orbital and $\langle p^\dagger_\downarrow p_\downarrow\rangle\sim 0.91$ for O $p_x$-orbital for the minority spin.
The un-quantized occupations of the former reflect considerable charge fluctuation $\langle d^\dagger_\downarrow p_\downarrow\rangle\sim 0.22$.
Upon \textit{slowly} increasing $U$ to 20 eV to suppress the charge fluctuation, $\langle d^\dagger_\downarrow p_\downarrow\rangle$ indeed \textit{smoothly} reduces to 0.14.
Correspondingly, $\langle d^\dagger_\downarrow d_\downarrow\rangle$ decreases to 0.09 and $\langle p^\dagger_\downarrow p_\downarrow\rangle$ increases to 0.96, indicating a well-quantized $|d^9p^6\rangle$ structure in agreement with the well-established ``charge transfer insulator''~\cite{Zaanen1985,Moskvin_PhysRevB_2011} characteristic of this material.

Note that since the dressed particles emerge at the scale of the dominant interaction $U$, as in the first example they generically constitute \textit{all} the lower-energy states.
Essentially, the lower-energy physics (such as long-range orders) simply form larger emergent objects using the emergent objects of higher energy as building blocks.
Therefore, the above `interaction annealing' procedure is immune to further emergence occurring at lower energy and insensitive to differences among the low-energy states.
Indeed, while having slightly higher energy, a ferromagnetic ground state can be stabilized under the realistic $U$ and yet slowly annealing $U$ to 20 eV gives very similar results and an identical $|d^9p^6\rangle$ quantized structure.

\subsection{5$d$ semi-metallic WTe$_2$}
For the above representative case of La$_2$CuO$_4$, an experienced researcher might be able to guess the obtained $|d^9p^6\rangle$ quantized structure without applying the proposed procedure, owing to the simplicity of the charge quantization in the remaining low-energy subspace.
Realistic functional materials, however, often contain much richer degrees of freedom to simply relying on experience-based guessing.

As a representative example, consider the semi-metallic material WTe$_2$~\cite{Lee2015, Das2019,Fei2018,Pankaj2019,Fangyuan_prepation, Ali2014, Soloyanov_Nat2015,Ma_Nat2019, Wang_npjcomp2019} with lattice structure shown in Fig.~\ref{WTe2_structure}.
Table~\ref{tab1} gives the local one-body density matrix $\rho_{nn'}\equiv\langle c_n^\dagger c_{n^\prime}\rangle$ of orbital $n$ for the DFT ground state under a reasonable value of $U=3$ eV~\cite{Kirchner2021, Linnartz2022} and the experimentally observed $T_d$ lattice structure~\cite{Lee2015}.
Since the intra-atomic interaction is generally not overwhelmingly large in $5d$ compounds, the result shows enormous charge fluctuation that makes it \textit{impossible} to identify a purely quantized valence and orbital structure desired for the physical understanding of low-energy dynamics.
(See Methods~\ref{inter_site_fluctuation} for quantification of inter-atomic charge fluctuation.)

Upon steadily suppressing the fluctuation via slowly `heating up' $U$ to 20 eV, one finds a much cleaner $\rho_{nn'}$ as in the lower half Tab.~\ref{tab1}.
In the $e_g^{\prime\prime}$ basis [c.f. Fig.~\ref{WTe2_structure}(c)], $\rho_{nn'}$ gives a well-quantized valence, orbital, and spin structure, corresponding to a \textit{dressed} $d^2$ orbital-polarized (OP2) structure shown in Fig.~\ref{WTe2_structure}(d).
One thus can reliably describe the electronic structure of the real material via an effective OP2 structure with dressed particles in the effective $\tilde{e}_g^{\prime\prime}$ orbitals (plus remaining fluctuation to $\ket{d^3\underline{L}}$ associated with the metallicity.)

Specifically, this energetically favored OP2 configuration contains double occupation of one of the symmetry-related dressed W $\tilde{e}_g^{\prime\prime}$ orbitals, as adiabatically connected to the one-body density matrix of a high-$U$ configuration shown in the lower part of Tab.~\ref{tab1}.
In the absence of long-range order, there would be another symmetry-related configuration with double occupation in the other dressed $\tilde{e}_g^{\prime\prime}$ orbitals.
Correspondingly, the low-energy dynamics must then contain that of an effective pseudo-spin-1/2 orbital fluctuation, or ``orbiton''~\cite{allen2001}.

Furthermore, our DFT ground contains a \textit{ferro-orbital ordering} of the dressed $\tilde{e}_{g2}^{\prime\prime}$ orbitals of W.
Such a ferro-orbital order of our identified OP2 structure offers a natural explanation for the experimentally observed octahedral deformation in $T_d$ lattice structure of this material below $\sim$550K of temperature~\cite{Tao_PhysRevB2020} and 6 GPa of pressure~\cite{Zhou_AIPadv2016}, since the lattice couples to the electronic charge and would follow the electronic ordering.

For consistency with experimental observations, it is easy to verify~\cite{supplementary} that the experimental observation of octahedral lattice distortion agrees much better with simple estimation using ionic radius of $d^2$ W ion than that via $d^3$ W ion~\cite{shannon1976revised}.
Similarly, the spin-0 structure of OP2 agrees perfectly with the experimentally observed diamagnetic response in WTe$_2$~\cite{Yang2021}.
These agreements with experiments confirm that the ferro-orbital ordered spin-0 picture obtained from our proposed `interaction annealing' procedure indeed captures the dominant higher-energy short-range correlation of WTe$_2$.
Such physical understandings are clearly not possible from the heavily fluctuating density matrix of the bare particles at $U=3$ eV.

\subsection{Application to competing electronic structures}\label{ATCES}

The proposed `interaction annealing' procedure is even more valuable for systems hosting competing local electronic structures that functionalizes modern materials.
Specifically, one can identify the potential emergent ionic configurations of well-defined valence and orbital structures using an artificially large $U$.
Furthermore, one can follow the adiabatic connection as $U$ reduces to the physical strength and identify the stable emergent ionic structure for each surviving configuration in real materials.

To give an example, consider a fictitious system of WTe$_2$ in an idealized $1T$ structure of higher symmetry shown in Fig.~\ref{WTe2_structure}(a) without the distortion of the $T_d$ structure.
As in many correlated materials, one finds multiple stable configurations, some of which may even appear extremely similar.
For example, the right columns of Tab~\ref{tab1} give $\rho_{nn^\prime}$ of two such stable configurations, containing charge fluctuation too strong to decipher the fully quantized emergent pictures even at a fictitious $U=8$ eV (upper part of Tab~\ref{tab1}), let alone to distinguish their corresponding physical properties and low-energy dynamics.

In spite of their strong similarity, however, their adiabatically connected interaction-annealed counterparts at $U=20$ eV (lower part of Tab~\ref{tab1}) display \textit{distinctive} quantized local electronic structures.
The former corresponds to an OP2 structure, while the latter contains fully occupied $e_g^{\prime\prime}$ orbitals, corresponding to the low-spin $d^4$ (LS4) configuration of W ion in Fig.~\ref{WTe2_structure}(d).
This LS4 configuration, while also diamagnetic, does not host the two-fold local orbital freedom of the OP2 configuration.
Therefore, it does not have low-energy ``orbiton'' dynamics, nor can it form a long-range orbital order.
This great contrast in physics nicely illustrates the value of our proposed procedure, especially when the fluctuation in real materials is strong enough to mask the essential physical distinction between stable configurations.

In fact, in functional materials, one expects multiple competing stable electronic structures that are sensitive to slightly different external conditions, such as temperature, pressure, or external fields~\cite{Kawai2018, Kushnireko2017, Hualei0516, Cao_TBG_2018}.
It is the qualitative distinct physical properties of these switchable stable configurations that grant the rich functionality of these materials.
Below, we proceed with this example to demonstrate that the `interaction annealing' procedure is not only valuable in deciphering the robust dressed objects, but also useful in identifying potential competing ones.

Exploring more thoroughly the interaction annealed ($U=20$ eV) case, one finds even more stable structures (fixed points), as shown in Fig.~\ref{WTe2_structure}(d), all with distinct well-defined ionic valence, orbital, and spin structures.
In addition to the OP2 and LS4 configuration, these also include a low-spin $d^2$ (LS2), a high-spin $d^3$ (HS3), and a orbital polarized $d^3$ (OP3) configurations.
One can also stabilize different spatial ordering of these ionic structures, such as ferro-orbital order (FO), antiferro-orbital order (AFO), ferromagnetic order (FM), and antiferromagnetic order (AFM).
This demonstrates that, indeed, interaction annealing is an efficient way to identify stable emergent electronic structures that compete to dictate the physical properties under various external conditions.

One can even use the interaction annealed systems to observe the competition between these stable structures, under a given external condition (the fixed lattice structure in this demonstration.)
As shown in Fig.~\ref{fig3}(a), these different stable structures naturally split into groups of distinct energy.
As the value of $U$ slowly `cools down', their physical quantities, such as total energy, \textit{smoothly} evolve.
Eventually, some of them become unstable and fall to other more stable ones, as indicated by the abrupt change in the total energy (and more in the density matrices), as emphasized in the inset of Fig~\ref{fig3}(a).
As illustrated in Fig~\ref{fig3}(b), such destabilization is associated with the vanishing of the corresponding fixed points manifesting as local energy minima in the phase space.
In functional materials, modulation of external conditions can alter the surviving local structures through varying the energy contour, thus enabling qualitative change of physical properties.

In this idealized structure with higher symmetry, all other configuration fall to the OP2 ones before $U$ is cooled to 5 eV~\cite{supplementary}.
That is, for real material the Rydberg-scale electronic correlation already develops a strong tendency toward local orbital polarization (and long-range ferro-orbital order at a lower energy scale).
The experimentally observed $T_d$ structure should, therefore, be considered driven by such electronic correlation, followed by further energy gain via lattice relaxation.

Using currently available many-body treatments, the proposed approach aims to reveal the dominant dressed objects that describe the slow dynamics of the system.
While this approach is not intended to improve these results (at least not by itself), as well demonstrated in the above examples, the obtained physical understanding is extremely valuable.

\section{Discussion}\label{discussion}

In summary, based on existing many-body computation methods, we propose a \textit{generic} `interaction annealing' approach to decipher the local electronic structure that dominates the low-energy dynamics.
We establish its \textit{robust} theoretical foundation and its applications using an exact treatment of a two-site Hubbard model and DFT calculations of two representative materials, $3d$ Mott insulator cuprate La${_2}$CuO${_4}$ and $5d$ semi-metal WTe${_2}$.
For the most fluctuating case of WTe$_2$, the obtained emergent local electronic structure makes possible an unprecedented explanation of several experimental observations~\cite{Lee2015, Das2019,Fei2018,Pankaj2019, Ali2014, Soloyanov_Nat2015,Ma_Nat2019, Wang_npjcomp2019}.
Finally, we demonstrate its effectiveness in studying competing local electronic structures in functional materials.
This approach is straightforward to implement in standard many-body computations for correlated materials.

\section{Methods}\label{methods}

\subsection{Obtaining low-energy effective Hamiltonian via numerical canonical transformation}\label{tllU}

Low-energy effective theory is one of the most effective means to capture and understand the key physics of a quantum system within a well-defined energy scale.
Conceptually, it can be rigorously constructed by integrating out the high-energy states in the path integral formulation.
Equivalently, it can also be derived by decoupling the remaining low-energy states from the high ones.
Specifically, one finds a unitary transformation\cite{chao1977kinetic,White2002}, $\mathcal{U}[\{c_i,c^{\dagger}_i\}]$, of the second quantized basis, $c_i$, spanning the one-body space indexed by $i$,
\begin{equation}
    \tilde{c_i}=\mathcal{U}^\dagger c_i\mathcal{U},
    \label{unit_basis}
\end{equation}
such that the Hamiltonian has a `block diagonal' form, $\Tilde{H}$, in the new `many-body dressed' representation, $\tilde{c_i}$ (freeing the low-energy subspace from the influence of the high-energy subspace),
\begin{equation}
    H[\{c_i,c^{\dagger}_i\}] = \Tilde{H}[\{\tilde{c}_i, \tilde{c}^\dagger_i\}].
    \label{eq2}
\end{equation}
Given
\begin{equation}
\begin{aligned}
    \Tilde{H}[\{\tilde{c}_i, \tilde{c}^\dagger_i\}] =& \Tilde{H}[\{\mathcal{U}^\dagger c_i\mathcal{U}, \mathcal{U}^\dagger c^\dagger_i \mathcal{U}\}]\\
    =& \mathcal{U}^\dagger \Tilde{H}[\{c_i,c^\dagger_i\}] \mathcal{U},
    \label{eq3}
\end{aligned}
\end{equation}
the unitary transformation $\mathcal{U}$ can thus be found by demanding
\begin{equation}
    \tilde{H}[\{c_i,c^\dagger_i\}] = \mathcal{U} H[\{c_i,c^\dagger_i\}] \mathcal{U}^\dagger,
    \label{eq4}
\end{equation}
to possess the desired block diagonal form.

As a simple illustration for such an emergent structure, consider a two-site Hubbard model in Eq.~\ref{HubboardH},
containing two electrons.
Let's first examine the emergent low-energy effective theories of this system in its correlated ($t \ll U$) regime.
In this regime, states with doubly occupied orbitals are the high-energy ones, as shown in the left panel of Fig~\ref{figs1}.
By defining the unitary transformation 
$\mathcal{U}=\exp(\frac{\theta}{2}\sum_\mu\frac{1}{2}(c^\dag_{1\mu}c_{2\mu}-c^\dag_{2\mu}c_{1\mu})(n_{2\bar{\mu}}-n_{1\bar{\mu}}))$ with $\theta=\arctan{(\frac{4t}{U})}$, 
the resulting effective Hamiltonian,
\begin{align}
\tilde{H}^{(t\ll U)}&=\left[\tilde{J}_{12} (\tilde{\mathbf{S}}_1\cdot\tilde{\mathbf{S}}_2-\frac{1}{4}\tilde{n}_1\tilde{n}_2)\right] 
\\&+\left[\tilde{U} (\tilde{c}^{\dagger}_{1\uparrow}\tilde{c}^{\dagger}_{1\downarrow}\tilde{c}_{1\downarrow}\tilde{c}_{1\uparrow}+ \tilde{c}^{\dagger}_{2\uparrow}\tilde{c}^{\dagger}_{2\downarrow}\tilde{c}_{2\downarrow}\tilde{c}_{2\uparrow}) + \frac{\tilde{J}_{12}}{2}(\tilde{c}^{\dagger}_{1\uparrow}\tilde{c}^{\dagger}_{1\downarrow}\tilde{c}_{2\downarrow}\tilde{c}_{2\uparrow}+ \tilde{c}^{\dagger}_{2\uparrow}\tilde{c}^{\dagger}_{2\downarrow}\tilde{c}_{1\downarrow}\tilde{c}_{1\uparrow})\right] \nonumber, 
\end{align}
cleanly separates the dynamics of the low-energy sector in the first line and high-energy one in the second.
The former no longer contains charge dynamics but only an emergent \textit{effective} inter-site anti-ferromagnetic super-exchange of strength $\tilde{J}_{12} = \frac{1}{2}(\sqrt{U^2+(4t)^2} - U)$, while the latter contains only effective inter-site charge dynamics of strength $\frac{\tilde{J}_{12}}{2}$ for the ``doublon'', $ \tilde{c}^{\dagger}_{i\uparrow}\tilde{c}^{\dagger}_{i\downarrow}$, two electrons tightly bound into a hard-core boson.
Here, $\tilde{\mathbf{S}}_i=\sum_{\mu\nu} \tilde{c}^\dag_{i\mu} \mathbf{\sigma}_{\mu\nu} \tilde{c}_{i\nu}$ denotes the \textit{effective} spin with Pauli matrices $\mathbf{\sigma}_{\mu\nu}$, $\tilde{n}_i = \tilde{c}^\dagger_{i\uparrow}\tilde{c}_{i\uparrow}+\tilde{c}^\dagger_{i\downarrow}\tilde{c}_{i\downarrow}$ the net \textit{dressed} particle number at site $i$, and $\tilde{U} = U+\frac{\tilde{J}_{12}}{2}$ the effective intra-site repulsion, or the effective potential energy of the doublon.
For clarity, Fig~\ref{figs1} gives the matrix presentation of the Hamiltonian in product states of the original $c^\dagger_i$ basis (left panel) and the dressed particles $\tilde{c}^\dagger_i$ (right panel).

One sees that the low-energy dynamics can \textit{always} be \textit{exactly} described by an effective theory containing only \textit{quantized} charge, orbital, and spin degrees of freedom, as the effective spin-$\frac{1}{2}$ shown here.

\subsection{The other (weakly correlated) stable fixed point}\label{tggU}

Concerning the highest-energy scale of the two-site Hubbard model, the other (weakly correlated) fixed point emerges for $t\gg U$.
In this regime, the (high-energy) rapid dynamics is now the inter-site kinetic process instead.
One thus first performs a canonical transformation via $\mathcal{V}_1 \equiv \exp\left(\frac{\pi}{4}~\sum_\mu\frac{1}{2}(c^\dagger_{2\mu}c_{1\mu}-c^\dagger_{1\mu}c_{2\mu})\right)$,
to decouple the energy of the bonding ($b^\dagger_\mu$) and anti-bonding ($a^\dagger_\mu$) orbitals at the one-body level,
\begin{align} 
\hat{H}^{(t\gg U)}&=t\sum_{\mu} (-b^\dagger_{\mu}b_{\mu}+a^\dagger_{\mu}a_{\mu}) \label{HtggU} \\
& + \frac{U}{2} \prod_\mu(b^\dagger_\mu b_\mu + a^\dagger_\mu  a_\mu) + \frac{U}{2} \prod_\mu(a^\dagger_\mu  b_\mu + b^\dagger_\mu a_\mu ), \nonumber
\end{align}
where $b^\dagger_\mu \equiv \mathcal{V}^\dagger_1 c^\dagger_{2\mu}\mathcal{V}_1 = \frac{1}{\sqrt{2}}(c^\dagger_{1\mu}+c^\dagger_{2\mu})$, and $a^\dagger_\mu \equiv \mathcal{V}^\dagger_1 c^\dagger_{1\mu} \mathcal{V}_1 = \frac{1}{\sqrt{2}}(c^\dagger_{1\mu}-c^\dagger_{2\mu})$ are the bonding and anti-bonding orbitals, respectively.

For an easier visualization, the left panel of Fig.~\ref{Intro_fig2} represents $H$ in the corresponding many-body basis.
One sees that states with the bonding and anti-bonding orbitals both singly occupied are perfectly decoupled from those with fully occupied bonding or anti-bonding orbitals.
However, the lowest-energy state with doubly occupied bonding orbitals and the highest-energy state with doubly occupied anti-bonding orbital are still coupled in the two-body level through the last term in Eq.~\ref{HtggU}.

We then perform another transformation via
$\mathcal{V}_2 \equiv \exp(\frac{\phi}{2}(a^\dag_{\uparrow}a^\dag_{\downarrow}b_{\downarrow}b_{\uparrow}-b^\dag_{\uparrow}b^\dag_{\downarrow}a_{\downarrow}a_{\uparrow}))$ with $\phi=\arctan{(U/4t)}$,
to further decouple them,
\begin{align}
\label{H_large_t}
\tilde{H}^{(t\gg U)}&=\left[-t \sum_{\mu} \tilde{b}^\dag_{\mu}\tilde{b}_{\mu}+(2t+2\tilde{\epsilon}_b)\tilde{b}^\dagger_\uparrow \tilde{b}^\dagger_\downarrow  \tilde{b}_\downarrow \tilde{b}_\uparrow\right]\\
&+\left[-\tilde{J}_{ba} (\tilde{\mathbf{S}}_b\cdot\tilde{\mathbf{S}}_a-\frac{1}{4}\tilde{n}_a\tilde{n}_b)\right] \nonumber \\
&+\left[t\sum_{\mu} \tilde{a}^\dag_{\mu}\tilde{a}_{\mu}+(-2t+2\tilde{\epsilon}_a) \tilde{a}^\dagger_\uparrow \tilde{a}^\dagger_\downarrow  \tilde{a}_\downarrow \tilde{a}_\uparrow \nonumber\right],
\end{align}
such that the slower dynamics is now cleanly separated into the lower (first line), middle (second line) and higher (third line) energy sectors, where $\tilde{\epsilon}_b = -\frac{1}{4}(\sqrt{(4t)^2+U^2}-U)$ and $\tilde{\epsilon}_a = \frac{1}{4}(\sqrt{(4t)^2+U^2}+U)$ denotes the emergent doublon potential energy for the \textit{effective} bonding and anti-boding orbitals.
The only remaining slow dynamics is now the emergent ferromagnetic inter-orbital Hund's coupling of strength $\tilde{J}_{ba}=U$ between the effective orbitals.
See Fig.~\ref{Intro_fig2} for the matrix representation of this canonical transformation.

As advocated in this manuscript, such an \textit{exact} emergent description for the slower dynamics is \textit{always} possible upon absorbing the rapid dynamics into the internal structure of the emergent objects, such as the effective bonding and anti-bonding orbitals,
\begin{align}
\tilde{b}^\dag_\mu
&= b^\dag_\mu-(\sin{\frac{\phi}{2}})~a^\dag_\mu a^\dag_{\bar{\mu}}b_{\bar{\mu}}+ (\cos{\frac{\phi}{2}}-1)~b^\dag_\mu b^\dag_{\bar{\mu}}b_{\bar{\mu}},\nonumber\\
\tilde{a}^\dag_\mu
&= a^\dag_\mu+(\sin{\frac{\phi}{2}})~b^\dag_\mu b^\dag_{\bar{\mu}}a_{\bar{\mu}}+(\cos{\frac{\phi}{2}}-1)~a^\dag_\mu a^\dag_{\bar{\mu}}a_{\bar{\mu}},
\label{b_tilde}
\end{align}
for this example.
Again, these emergent objects contain fully \textit{quantized} effective charge, orbital, and spin degrees of freedom.

Notice that such energy separation and the residual dynamics is again adiabatically connected to the $t\gg U$ fixed point.
Indeed, boosting up the dominant kinetic process, or more conveniently reducing $\frac{U}{t}$ toward $0^+$, one finds that the main structures of Eq.~\ref{H_large_t} and \ref{b_tilde} remain qualitatively intact, with only quantitative change in the renormalized coefficients.
Therefore, the emergent objects share the characteristics of those with negligible intra-atomic interaction.
One can thus apply the ``interaction annealing'' procedure, but toward \textit{smaller} $U$, to access the characteristics of emergent objects belong to this weakly correlated fixed point.
This nicely explains the great success in application of LDA Wannier orbitals within the DFT framework: their qualitative characteristics represent those of the emergent objects in weakly correlated systems, even those with weak intra-atomic interactions.

\subsection{Computational details of density functional calculation}
Our first-principles calculations were performed using density functional theory (DFT)-based~\cite{DFT1, DFT2} {\sc Wien2k} package~\cite{WienBlaha} with the spin-polarized LDA+$U$~\cite{Anisimov1993, Liechtenstein1995} and the linearized augmented plane wave (LAPW)~\cite{Singh} implementation~\cite{Blaha1990}.
Monkhorst-Pack $k$-point grids for Brillouin zone sampling are set to $8\times15\times3$ for bulk WTe$_2$ with both $1T$ and $T_d$ structures.
WTe$_2$ crystallizes in an orthorhombic Bravais lattice, with a $T_d$ structure and space group $Pmn21$ (No. 31).
We used the experimental lattice constants reported by Brown \textit{et al}.~\cite{Brown1966TheCS}, $a=6.28\mathring{A}$, $b=3.50\mathring{A}$, $c=14.07\mathring{A}$, which were also confirmed by other experiments~\cite{LeeChiaHui2015}.
As the higher-symmetry $1T$ structure has not been observed in experiments, we built the structure with the same interlayer thickness as the experimental $T_d$ structure and optimized the intralayer lattice parameters through atomic relaxation.
The lattice constants used for calculations in this work are $a=6.06\mathring{A}$, $b=3.50\mathring{A}$, $c=14.07\mathring{A}$, with the same space group $Pmn21$ (No. 31).

In the interaction annealing process, we set the effective $U$ on W $d$-orbitals within the range of $3-20$ eV. 
The `heat up' and `cool down' processes correspond to smoothly increasing and decreasing the interaction strength $U$ in 0.2 eV and 0.1 eV per step with a mix factor of 0.05.

\subsection{One-body density matrix $\rho_{nn^\prime}$ among the local W $t_{2g}$ orbitals}

At the high-$U$ ($\sim$ 20 eV) regime, one finds many stable structures in the LDA+$U$ calculations, all with distinct well-defined ionic valence, orbital, and spin structures.
In order to reveal the potential spontaneous symmetry breaking of the system, the idealized 1$T$ structure of higher symmetry with doubled W in one unit cell is thus built.
Table \ref{tabs1} lists the one-body density matrix $\rho_{nn^\prime}$ among the local $t_{2g}$ orbitals of two W atoms with index $n$ for different $U \sim 20$ eV configurations in 1$T$ structure.

\subsection{Illustration of competing configurations}

Figure~\ref{detail_annealing} shows the detail information of competing structures under the symmetric $1T$ structure of WTe$_2$ and its adiabatic connection upon `decreasing' the interaction strength $U$.
Through a smooth evolution of total energy (for 4 chemical formula units) upon reduction of $U$ from the large limit $\sim20$ eV to the realistic value $\sim3$ eV, the total energy evolves smoothly until some configurations become unstable and fall to other more stable ones, as indicated by the abrupt `jump' in the total energy curves.
All other configurations fall to the OP2 ones before $U$ is cooled to 5 eV. 

\subsection{Stable configuration via electron-lattice coupling in $T_d$ structure}

With the help of lattice distortion towards lower symmetry, which is consistent with the symmetry-broken electronic state, this electronic configuration would be further stabilized.
Fig.~\ref{Td_annealing} shows the stable configuration of WTe$_2$ and its adiabatic connection upon `heating up' the interaction strength $U$.
As $U$ slowly increases from the realistic value $3$ eV to large limit $20$ eV, the configuration is OP2 with ferro orbital order, corresponding to the two electrons both occupying one of the two degenerate orbitals $eg''_2$ with long-range ferro-orbital order.

\subsection{Strong reduction of inter-atomic charge fluctuation}
\label{inter_site_fluctuation}

The strong charge fluctuation (and its suppression by intra-atomic repulsion $U$) is reflected in difficulty in deciphering the dominant characteristics of the emergent quantized objects.
Microscopically, such charge fluctuation is more directly reflected in the inter-atomic one-body density matrix $\rho$.
To quantitatively demonstrate the effect of charge fluctuations and their suppression at larger $U$, Tab~\ref{tabs2} gives a few representative elements of $\rho_{pd}$ between Te $p$-orbitals and W $d$-orbitals, which are computed via symmetry respecting atomic Wannier orbitals~\cite{Wei2002}.

Consider for example the first row.
At realistic $U \sim 3$ eV, $\frac{\rho_{pd}}{\rho_{pp}-\rho_{dd}} \sim 1.33 >$ 1, indicating a strong inter-atomic charge fluctuation between these two orbitals.
In contrast, under the increased $U \sim 20$ eV, $\frac{\rho_{pd}}{\rho_{pp}-\rho_{dd}} \sim 0.29 \ll$ 1, reflecting a dramatic suppression of the inter-atomic charge fluctuation, which makes it much easier to decipher the low-energy emergent objects.

\subsection{Estimated octahedral distortion from ionic radius~\cite{shannon1976revised} for $d^2$ and $d^3$ configurations  in WTe$_2$ with $T_{d}$ structure}

Tab~\ref{tab:distortion_struct} gives the calculated average deviations of both W-Te bond length, $\Delta=\sum_{i=1}^6\left|d_i-d_{\text{mean}}\right|$, and the Te-W-Te intersection angle $\phi_i$, $\Sigma=\sum_{i=1}^{12}\left|90^\circ-\phi_i\right|$, of the local WTe$_6$ octahedron.
Evidently, compared with W$^{3+}$Te$^{1.5-}_2$ ($d^3$), W$^{4+}$Te$^{2-}_2$ ($d^2$) gives a superior agreement with the experimental structure.
In fact, the negligible $\Delta$ for W$^{3+}$Te$^{1.5-}_2$ indicates that W$^{3+}$ ion is too large to be compatible with the $T_d$ structure. 

\section{Declaration statements}\label{statements}

\subsection{Data Availability}

The data that support the findings of this study are available from the corresponding author upon reasonable request.

\subsection{Acknowledgements}

This work is supported by the National Natural Science Foundation of China (NSFC) under Grants No. 12274287 and No. 12042507 and the Innovation Program for Quantum Science and Technology No. 2021ZD0301900.
R.J. acknowledges additional support by UKRI Future Leaders Fellowship [MR/V023926/1].

\subsection{Author Contributions}

R.J. conducted the exact many-body treatment of the two-site Hubbard model. 
R.J. and F.G. carried out and analyzed the DFT calculations.
W.K. supervised the project. 
All authors contributed to the conceptualization of the project, methodology design, discussion of the results, and the writing of the manuscript.

\subsection{Competing Interests}

The authors declare no competing interests.

\bibliography{Manuscript}

@book{Robert2001,
  author    = {Robert W. Cahn},
  title     = {Functional Materials},
  booktitle = {The Coming of Materials Science},
  publisher = {Pergamon},
  address = {Amsterdam},
  publisher = {New York},
  volume    = {5},
  pages     = {253-304},
  year      = {2001}
}

@article{Schrieffer1966,
  author = {Schrieffer, J. R. and Wolff, P. A.},
  title = {Relation between the Anderson and Kondo Hamiltonians},
  journal = {Phys. Rev.},
  volume = {149},
  pages = {491--492},
  year = {1966}
}

@article{Zaanen1988,
  author = {Zaanen, Jan and Ole\'{s}, Andrzej M.},
  title = {Canonical perturbation theory and the two-band model for high-${T}_{c}$ superconductors},
  journal = {Phys. Rev. B},
  volume = {37},
  pages = {9423--9438},
  year = {1988}
}

@article{Wei2002,
  author = {Ku, Wei and Rosner, H. and Pickett, W. E. and Scalettar, R. T.},
  title = {Insulating Ferromagnetism in $\rm{La_4Ba_2Cu_2O_{10}}$: An Ab Initio Wannier Function Analysis},
  journal = {Phys. Rev. Lett.},
  volume = {89},
  pages = {167204},
  year = {2002},
}

@article{Kanungo2014,
  author = {Kanungo, Sudipta and Yan, Binghai and Jansen, Martin and Felser, Claudia},
  title = {Ab-inito study of low temperature magnetic properties of double perovskite $\rm{Sr_2FeOsO_6}$},
  journal = {Phys. Rev. B},
  volume = {89},
  pages = {214414},
  year = {2014}
}

@article{Kato1950,
  author = {Kato ,Tosio},
  title = {On the Adiabatic Theorem of Quantum Mechanics},
  journal = {J. Phys. Soc. Jpn.},
  volume = {5},
  pages = {435-439},
  year = {1950}
}

@book{Sahu2019,
  author = {Dipti Sahu},
  title = {Functional Materials},
  publisher = {IntechOpen},
  address = {London},
  year = {2019},
}

@book{Sakurai1994,
  author        = "Sakurai, Jun John",
  title         = "{Modern quantum mechanics; rev. ed.}",
  publisher     = "Addison-Wesley",
  address       = "Reading, MA",
  year          = "1994"
}

@article{Hualei0516,
  author={Hualei Sun and Mengwu Huo and Xunwu Hu and Jingyuan Li and Yifeng Han and Lingyun Tang and Zhongquan Mao and Pengtao Yang and Bosen Wang and Jinguang Cheng and Dao-Xin Yao and Guang-Ming Zhang and Meng Wang},
  title={Superconductivity near 80 Kelvin in single crystals of $\mathrm{La_3Ni_2O_7}$ under pressure}, 
  journal = {Nature},
  volume = {621},
  pages = {493-498},
  year={2023}
}

@article{Arovas2022,
   author={Arovas, Daniel P. and Berg, Erez and Kivelson, Steven A. and Raghu, Srinivas},
   title={The Hubbard Model},
   journal={Annu. Rev. Condens. Matter Phys.},
   volume={13},
   pages={239-274},
   number={1},
   year={2022}
}

@article{chao1977kinetic,
  author={Chao, KA and Spalek, J and Oles, AM},
  title={Kinetic exchange interaction in a narrow S-band},
  journal={J. Phys. C: Solid State Phys.},
  volume={10},
  pages={L271},
  number={10},
  year={1977}
}

@article{White2002,
  author = {White, Steven},
  title = {Numerical canonical transformation approach to quantum many-body problems},
  journal = {J. Chem. Phys.},
  volume = {117},
  pages={7472–7482},
  year = {2002}
}

@article{DFT1,
  author = {Hohenberg, P. and Kohn, W.},
  title = {Inhomogeneous Electron Gas},
  journal = {Phys. Rev.},
  volume = {136},
  pages = {B864--B871},
  year = {1964}
}

@article{DFT2,
  author = {Kohn, W. and Sham, L. J.},
  title = {Self-Consistent Equations Including Exchange and Correlation Effects},
  journal = {Phys. Rev.},
  volume = {140},
  pages = {A1133--A1138},
  year = {1965}
}

@article{Anderson1972,
  author = {P. W. Anderson},
  title = {More Is Different},
  journal = {Science},
  volume = {177},
  number = {4047},
  pages = {393-396},
  year = {1972}
}

@article {Dagotto2005,
  author = {Dagotto, Elbio},
  title = {Complexity in Strongly Correlated Electronic Systems},
  journal = {Science},
  volume = {309},
  number = {5732},
  pages = {257--262},
  year = {2005}
}

@article{Dagotto1994,
  author = {Dagotto, Elbio},
  title = {Correlated electrons in high-temperature superconductors},
  journal = {Rev. Mod. Phys.},
  volume = {66},
  issue = {3},
  pages = {763--840},
  numpages = {0},
  year = {1994}
}

@article{Yang2021,
  author = {Yang, Li and Wu, Hao and Zhang, Liang and Zhang, Wenfeng and Li, Luying and Kawakami, Tappei and Sugawara, Katsuaki and Sato, Takafumi and Zhang, Gaojie and Gao, Pengfei and Muhammad, Younis and Wen, Xiaokun and Tao, Boran and Guo, Fei and Chang, Haixin},
  title = {Highly Tunable Near-Room Temperature Ferromagnetism in \text{Cr}-Doped Layered ${T}_{d}$-$\rm{WTe_2}$},
  journal = {Adv. Funct. Mater.},
  volume = {31},
  pages = {2008116},
  year = {2021}
}

@article{Fei2018,
  author = {Fei, Zaiyao and Zhao, Wenjin and Palomaki, Tauno and Sun, Bosong and Miller, Moira and Zhao, Zhiying and Yan, J.-Q and Xu, Xiaodong and Cobden, David},
  title = {Ferroelectric switching of a two-dimensional metal},
  journal = {Nature},
  volume = {560},
  pages = {pages336–339},
  year = {2018}
}

@article{Das2019,
  author = {P K Das and D Di Sante and F Cilento and C Bigi and D Kopic and D Soranzio and A Sterzi and J A Krieger and I Vobornik and J Fujii and T Okuda and V N Strocov and M B H Breese and F Parmigiani and G Rossi and S Picozzi and R Thomale and G Sangiovanni and R J Cava and G Panaccione},
  title = {Electronic properties of candidate type-\text{II} Weyl semimetal $\rm{WTe_2}$. A review perspective},
  journal = {Electron. Struct.},
  volume = {1},
  pages = {014003},
  year = {2019}
}

@article{Kawai2018,
  author = {Kawai, M and Nabeshima, F and Maeda, A},
  title = {Transport properties of {FeSe} epitaxial thin films under in-plane strain},
  journal = {J. Phys.: Conf. Ser.},
  pages = {012023},
  volume = {1054},
  year = {2018}
}

@article{Kushnireko2017,
  author = {Kushnirenko, Y. S. and Kordyuk, A. A. and Fedorov, A. V. and Haubold, E. and Wolf, T. and B\"uchner, B. and Borisenko, S. V.},
  title = {Anomalous temperature evolution of the electronic structure of {FeSe}},
  journal = {Phys. Rev. B},
  volume = {96},
  issue = {10},
  pages = {100504(R)},
  numpages = {5},
  year = {2017}
}

@article{Pankaj2019,
  author = {Pankaj Sharma  and Fei-Xiang Xiang  and Ding-Fu Shao  and Dawei Zhang  and Evgeny Y. Tsymbal  and Alex R. Hamilton  and Jan Seidel },
  title = {A room-temperature ferroelectric semimetal},
  journal = {Sci. Adv.},
  volume = {5},
  number = {7},
  pages = {eaax5080},
  year = {2019}
}

@article{Ali2014,
  author = {Ali, {Mazhar N.} and Jun Xiong and Steven Flynn and Jing Tao and Gibson, {Quinn D.} and Schoop, {Leslie M.} and Tian Liang and Neel Haldolaarachchige and Max Hirschberger and Ong, {N. P.} and Cava, {R. J.}},
  title = {Large, non-saturating magnetoresistance in $\rm{WTe_2}$},
  journal = {Nature},
  volume = {514},
  pages = {205--208},
  year = {2014}
}

@article{Anisimov1993,
  author = {Anisimov, V. I. and Solovyev, I. V. and Korotin, M. A. and Czy\.{z}yk, M. T. and Sawatzky, G. A.},
  title = {Density-functional theory and \text{NiO} photoemission spectra},
  journal = {Phys. Rev. B},
  volume = {48},
  issue = {23},
  pages = {16929--16934},
  numpages = {0},
  year = {1993}
}

@article{Liechtenstein1995,
  title = {Density-functional theory and strong interactions: Orbital ordering in \text{M}ott-Hubbard insulators},
  author = {Liechtenstein, A. I. and Anisimov, V. I. and Zaanen, J.},
  journal = {Phys. Rev. B},
  volume = {52},
  issue = {8},
  pages = {R5467--R5470},
  numpages = {0},
  year = {1995}
}

@article{Fangyuan_prepation,
  author={Fangyuan Gu and Ruoshi Jiang and Wei Ku},
  title={2{D} ferroelectricity accompanying antiferro-orbital order in semi-metallic {WTe$_2$}}, 
  journal={arXiv},
  doi={https://arxiv.org/abs/2507.18438},
  year={2025}
}

@article{Lee2015,
  author = {{Lee}, Chia-Hui and {Silva}, Eduardo Cruz and {Calderin}, Lazaro and {Nguyen}, Minh An T. and {Hollander}, Matthew J. and {Bersch}, Brian and {Mallouk}, Thomas E. and {Robinson}, Joshua A.},
  title = {Tungsten Ditelluride: a layered semimetal},
  journal = {Sci. Rep.},
  volume = {5},
  pages = {10013},
  year = {2015}
}

@misc{Jiang_int_valence,
    note={Ruoshi Jiang, Xiang Li, Xinyao Zhang and Wei Ku, in preparation}
}

@misc{supplementary,
    note={See Methods for details.}
}

@article{Jiang2024,
  author = {Jiang, Ruoshi and Hou, Jinning and Fan, Zhiyu and Lang, Zi-Jian and Ku, Wei},
  title = {Pressure Driven Fractionalization of Ionic Spins Results in Cupratelike High-${T}_{c}$ Superconductivity in $\rm{La_3Ni_2O_7}$},
  journal = {Phys. Rev. Lett.},
  volume = {132},
  pages = {126503},
  numpages = {7},
  year = {2024}
}

@article{Cao_TBG_2018,
  author={Yuan Cao and Valla Fatemi and Shiang Fang and Kenji Watanabe and Takashi Taniguchi and Efthimios Kaxiras and Pablo Jarillo-Herrero},
  title={Unconventional superconductivity in magic-angle graphene superlattices},
  journal={Nature},
  volume={556},
  pages={43-50},
  year={2018}
}

@article{Hubbard1963Electron,
  author={J. Hubbard},
  title={Electron correlations in narrow energy bands},
  journal={Proc. R. Soc. Lond. A},
  volume={276},
  pages={238-257},
  year={1963}
}

@article{Anisimov1997,
  author = {Vladimir I Anisimov and F Aryasetiawan and A I Lichtenstein},
  title = {First-principles calculations of the electronic structure and spectra of strongly correlated systems: the $\rm{LDA+\textit{U}}$ method},
  journal = {J. Phys.: Condens. Matter},
  volume = {9},
  number = {4},
	 pages = {767--808},
  year = {1997}
}

@article{Tao_PhysRevB2020,
  author = {Tao, Yu and Schneeloch, John A. and Aczel, Adam A. and Louca, Despina},
  title = {${T}_{d}$ to $1{T}^{\ensuremath{'}}$ structural phase transition in the $\rm{WTe_2}$ Weyl semimetal},
  journal = {Phys. Rev. B},
  volume = {102},
  pages = {060103},
  numpages = {5},
  year = {2020}
}

@article{shannon1976revised,
  author={Shannon, Robert D},
  title={Revised effective ionic radii and systematic studies of interatomic distances in halides and chalcogenides},
  journal={Acta Crystallogr. A},
  volume={32},
  number={5},
  pages={751--767},
  year={1976}
}

@article{Soloyanov_Nat2015,
  author = {Soluyanov, Alexey A. and Gresch, Dominik and Wang, Zhijun and Wu, QuanSheng and Troyer, Matthias and Dai, Xi and Bernevig, B. Andrei},
  title = {Type-\text{II} Weyl semimetals},
  journal = {Nature},
  volume = {527},
  number = {7579},
  pages = {495--498},
  year = {2015}
}

@article{Ma_Nat2019,
  author = {Ma, Qiong and Xu, Su-Yang and Shen, Huitao and MacNeill, David and Fatemi, Valla and Chang, Tay-Rong and Mier Valdivia, Andr{\'e}s M. and Wu, Sanfeng and Du, Zongzheng and Hsu, Chuang-Han and Fang, Shiang and Gibson, Quinn D. and Watanabe, Kenji and Taniguchi, Takashi and Cava, Robert J. and Kaxiras, Efthimios and Lu, Hai-Zhou and Lin, Hsin and Fu, Liang and Gedik, Nuh and Jarillo-Herrero, Pablo},
  title = {Observation of the nonlinear Hall effect under time-reversal-symmetric conditions},
  journal = {Nature},
  number = {7739},
  pages = {337--342},
  volume = {565},
  year = {2019}
}

@article{Wang_npjcomp2019,
  author = {Wang, Hua and Qian, Xiaofeng},
  title = {Ferroelectric nonlinear anomalous Hall effect in few-layer $\rm{WTe_2}$},
  journal = {npj Comput. Mater.},
  number = {1},
  pages = {119},
  volume = {5},
  year = {2019}
}

@article{Zhou_AIPadv2016,
  author = {Zhou, Yonghui and Chen, Xuliang and Li, Nana and Zhang, Ranran and Wang, Xuefei and An, Chao and Zhou, Ying and Pan, Xingchen and Song, Fengqi and Wang, Baigeng and Yang, Wenge and Yang, Zhaorong and Zhang, Yuheng},
  title = {Pressure-induced Td to 1T' structural phase transition in $\rm{WTe_2}$},
  journal = {AIP Adv.},
  volume = {6},
  number = {7},
  pages = {075008},
  year = {2016}
}

@book{granas2003fixed,
  title={Fixed point theory},
  author={Granas, Andrzej and Dugundji, James},
  month={06},
  volume={14},
  page={690},
  issn={1439-7382},
  journal={Springer},
  year={2003}
}

@book{Nigel1992,
  author={Goldenfeld, N.},
  title={Lectures On Phase Transitions And The Renormalization Group},
  journal={CRC Press.},
  year={1992}
}

@article{Kirchner2021,
  author= {Kirchner-Hall, Nicole E. and Zhao, Wayne and Xiong, Yihuang and Timrov, Iurii and Dabo, Ismaila},
  title = {Extensive Benchmarking of $\rm{LDA+\textit{U}}$ Calculations for Predicting Band Gaps},
  journal = {Appl. Sci.},
  volume = {11},
  year = {2021}
}

@article{QuantumAlgebra,
	author  = {Johannes Feist and contributors},
	title   = {QuantumAlgebra.jl},
	doi   = {https://github.com/jfeist/QuantumAlgebra.jl},
	version = {v1.1.0},
	year    = {2021}
}

@article{Sanchez-Barquilla2020,
  author = {{S{\'a}nchez-Barquilla}, M. and Silva, R. E. F. and Feist, J.},
  title = {Cumulant Expansion for the Treatment of Light-Matter Interactions in Arbitrary Material Structures},
  journal = {J. Chem. Phys.},
  volume = {152},
  pages = {034108},
  number = {3},
  year = {2020}
}

@article{spherharm_web,
  author = {Qian, Zhang and Khaldoon, Ghanem},
  title = {Spherical Harmonics Visualization},
  doi = {https://www.cond-mat.de/teaching/QM/JSim/spherharm.html},
  year = {2013}
}

@article{PIE1,
  author = {Stevanovi\'{c}, Vladan and d'Avezac, Mayeul and Zunger, Alex},
  title = {Simple Point-Ion Electrostatic Model Explains the Cation Distribution in Spinel Oxides},
  journal = {Phys. Rev. Lett.},
  volume = {105},
  issue = {7},
  pages = {075501},
  numpages = {4},
  year = {2010}
}

@article{PIE2,
  author = {Stevanovi\'{c}, Vladan and d'Avezac, Mayeul and Zunger, Alex},
  title = {Universal Electrostatic Origin of Cation Ordering in $\rm{A_2BO_4}$ Spinel Oxides},
  journal = {J. Am. Chem. Soc.},
  volume = {133},
  number = {30},
  pages = {11649-11654},
  year = {2011}
}

@article{electrostatic,
  title = {Electrostatic Model of Atomic Ordering in Complex Perovskite Alloys},
  author = {Bellaiche, L. and Vanderbilt, David},
  journal = {Phys. Rev. Lett.},
  volume = {81},
  issue = {6},
  pages = {1318--1321},
  numpages = {0},
  year = {1998}
}

@article{Chargetrans,
  author = {Wu, Zhigang and Krakauer, Henry},
  title = "{Charge transfer model of atomic ordering in complex perovskite alloys}",
  journal = {AIP Conf. Proc.},
  volume = {535},
  number = {1},
  pages = {121-128},
  year = {2000}
}

@article{Moskvin_PhysRevB_2011,
  title = {True charge-transfer gap in parent insulating cuprates},
  author = {Moskvin, A. S.},
  journal = {Phys. Rev. B},
  volume = {84},
  issue = {7},
  pages = {075116},
  numpages = {11},
  year = {2011}
}

@article{Nelson2015,
  title = {What is the Valence of Mn in $\mathrm{Ga_{1-x}Mn_xN}$?},
  author = {Nelson, Ryky and Berlijn, Tom and Moreno, Juana and Jarrell, Mark and Ku, Wei},
  journal = {Phys. Rev. Lett.},
  volume = {115},
  issue = {19},
  pages = {197203},
  numpages = {6},
  year = {2015}
}

@article{Taillefer2010,
  author = {Taillefer, Louis},
  title = {Scattering and Pairing in Cuprate Superconductors},
  journal = {Annu. Rev. Condens. Matter Phys.},
  volume = {1},
  pages={51–70},
  year = {2010}
}

@article{yokoyama1987variational,
  title={Variational Monte-Carlo studies of Hubbard model. i},
  author={Yokoyama, Hisatoshi and Shiba, Hiroyuki},
  journal={J. Phys. Soc. Jpn.},
  volume={56},
  number={4},
  pages={1490--1506},
  year={1987}
}

@article{schollwock2005density,
  title={The density-matrix renormalization group},
  author={Schollw{\"o}ck, Ulrich},
  journal={Rev. Mod. Phys.},
  volume={77},
  number={1},
  pages={259--315},
  year={2005}
}

@article{evenbly2011tensor,
  title={Tensor network states and geometry},
  author={Evenbly, Glen and Vidal, Guifr{\'e}},
  journal={J. Stat. Phys.},
  volume={145},
  pages={891--918},
  year={2011}
}

@article{Anisimov2004,
  title = {Computation of stripes in cuprates within the $\rm{LDA+\textit{U}}$ method},
  author = {Anisimov, V. I. and Korotin, M. A. and Mylnikova, A. S. and Kozhevnikov, A. V. and Korotin, Dm. M. and Lorenzana, J.},
  journal = {Phys. Rev. B},
  volume = {70},
  issue = {17},
  pages = {172501},
  numpages = {4},
  year = {2004}
}

@article{Zaanen1985,
  title = {Band gaps and electronic structure of transition-metal compounds},
  author = {Zaanen, J. and Sawatzky, G. A. and Allen, J. W.},
  journal = {Phys. Rev. Lett.},
  volume = {55},
  issue = {4},
  pages = {418--421},
  numpages = {0},
  year = {1985}
}

@article{Linnartz2022,
  title = {Fermi surface and nested magnetic breakdown in $\rm{WTe_2}$},
  author = {Linnartz, J. F. and M\"uller, C. S. A. and Hsu, Yu-Te and Nielsen, C. Breth and Bremholm, M. and Hussey, N. E. and Carrington, A. and Wiedmann, S.},
  journal = {Phys. Rev. Res.},
  volume = {4},
  issue = {1},
  pages = {L012005},
  numpages = {5},
  year = {2022}
}

@article{allen2001,
  title={First glimpse of the orbiton},
  author={Allen, Philip B and Perebeinos, Vasili},
  journal={Nature},
  volume={410},
  number={6825},
  pages={155--158},
  year={2001}
}

@Book{Singh,
  author = {David J. Singh},
  title = {Planewaves, Pseudopotentials and the LAPW Method},
  publisher = {Springer New York, NY},
  year = {2006},
  address = {New York},
}

@article{Blaha1990,
  author = {P. Blaha and K. Schwarz and P. Sorantin and S.B. Trickey},
  title = {Full-potential, linearized augmented plane wave programs for crystalline systems},
  journal = {Comput. Phys. Commun.},
  volume = {59},
  number = {2},
  pages = {399-415},
  year = {1990}
}

@article{LeeChiaHui2015,
  author = {{Lee}, Chia-Hui and {Silva}, Eduardo Cruz and {Calderin}, Lazaro and {Nguyen}, Minh An T. and {Hollander}, Matthew J. and {Bersch}, Brian and {Mallouk}, Thomas E. and {Robinson}, Joshua A.},
  title = "{Tungsten Ditelluride: a layered semimetal}",
  journal = {Sci. Rep.},
  volume = {5},
  pages = {10013},
  year = {2015}
}

@article{Brown1966TheCS,
  title={The crystal structures of $\rm{WTe_2}$ and high‐temperature $\rm{MoTe_2}$},
  author={B. E. Brown},
  journal={Acta Crystallogr.},
  volume={20},
  pages={268-274},
  year={1966}
}

@article{WienBlaha,
  author = {Blaha, Peter and Schwarz, Karlheinz and Tran, Fabien and Laskowski, Robert and Madsen, Georg K. H. and Marks, Laurence D.},
  title = "{WIEN2k: An APW+lo program for calculating the properties of solids}",
  journal = {J. Chem. Phys.},
  volume = {152},
  number = {7},
  pages = {074101},
  year = {2020}
}

@article{Larson2007Nonresonant,
  title = {Nonresonant Inelastic X-Ray Scattering and Energy-Resolved Wannier Function Investigation of $d\mathrm{\text{\ensuremath{-}}}d$ Excitations in \text{NiO} and \text{CoO}},
  author = {Larson, B. C. and Ku, Wei and Tischler, J. Z. and Lee, Chi-Cheng and Restrepo, O. D. and Eguiluz, A. G. and Zschack, P. and Finkelstein, K. D.},
  journal = {Phys. Rev. Lett.},
  volume = {99},
  issue = {2},
  pages = {026401},
  numpages = {4},
  year = {2007}
}

@article{Walters2009Effect,
  author = {Walters, Andrew and Perring, Tg and Caux, Jean-S|[eacute]|bastien and Savici, A. and Gu, Genda and Lee, Chi-Cheng and Ku, Wei and Zaliznyak, I.},
  month = {10},
  pages = {867-872},
  title = {Effect of covalent bonding on magnetism and the missing neutron intensity in copper oxide compounds},
  volume = {5},
  journal = {Nat. Phys.},
  year = {2009}
}

@article{Peter2008Dynamical,
  author = {Peter Abbamonte  and Tim Graber  and James P. Reed  and Serban Smadici  and Chen-Lin Yeh  and Abhay Shukla  and Jean-Pascal Rueff  and Wei Ku },
  title = {Dynamical reconstruction of the exciton in \text{LiF} with inelastic x-ray scattering},
  journal = {Proc. Natl. Acad. Sci. U.S.A.},
  volume = {105},
  number = {34},
  pages = {12159-12163},
  year = {2008}
}

@article{Pernal2018Electron,
  title = {Electron Correlation from the Adiabatic Connection for Multireference Wave Functions},
  author = {Pernal, Katarzyna},
  journal = {Phys. Rev. Lett.},
  volume = {120},
  issue = {1},
  pages = {013001},
  numpages = {5},
  year = {2018}
}

@article{Eshuis2012Electron,
  author = {Eshuis, Henk and Bates, Jefferson and Furche, Filipp},
  title = {Electron correlation methods based on the random phase approximation},
  journal = {Theor. Chem. Acc.},
  volume = {131},
  pages = {1-18},
  year = {2012}
}

@article{jiang2025Emergence,
  author={Ruoshi Jiang and Zhiyu Fan and Bartomeu Monserrat and Wei Ku},
  title={Pressure-induced trans-proximate correlation in {La$_4$Ni$_3$O$_{10}$} and possible routes to enhance its superconductivity}, 
  journal={arXiv},
  doi={https://arxiv.org/abs/2505.13442}, 
  year={2025}
}

\clearpage

\begin{figure}[h!]
\centering
\includegraphics[width=0.8\columnwidth]{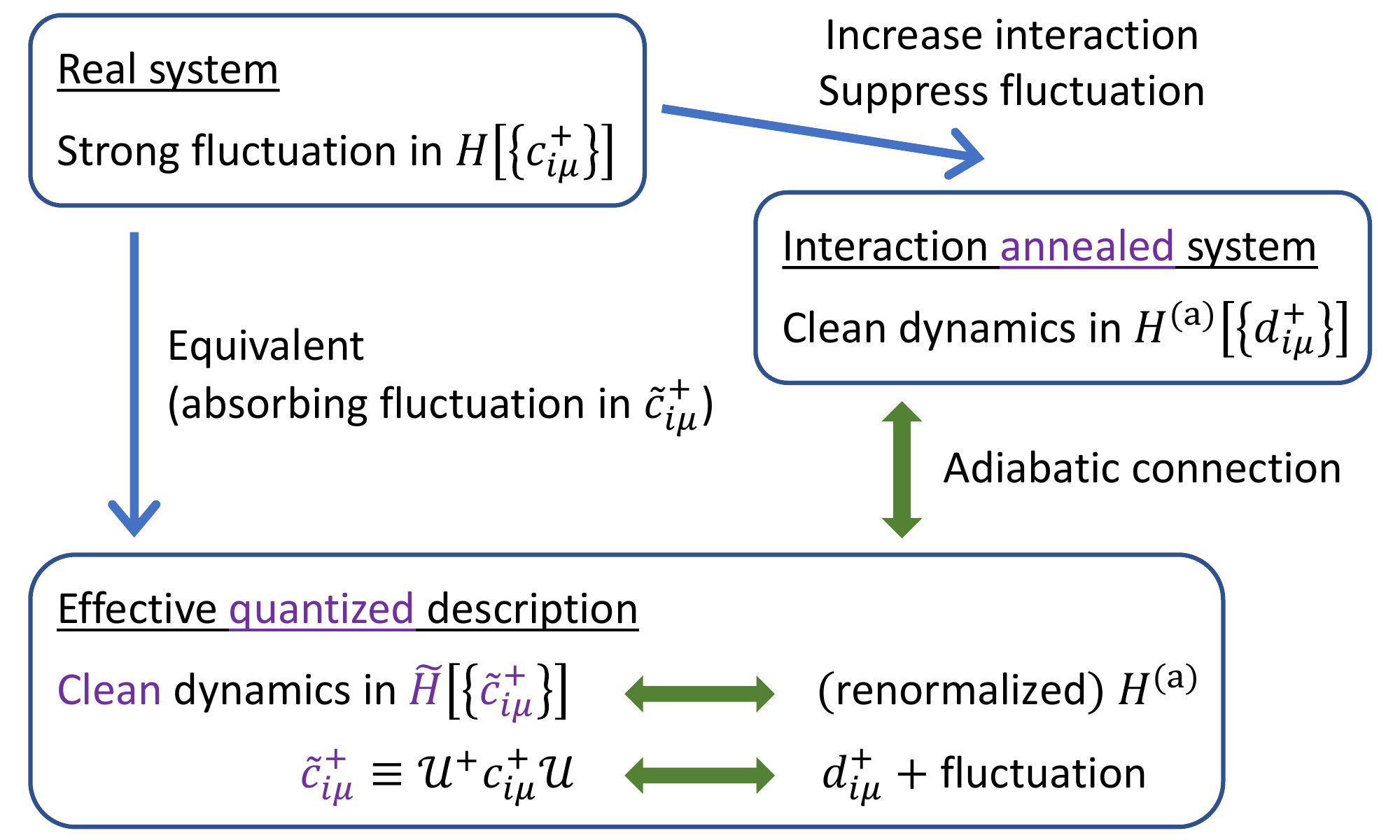}
\vspace{0.5cm}
\caption{Illustration on obtaining effective \textit{quantized} description through the interaction annealing procedure.
Realistic Hamiltonian $H$ for bare particle $c_i^\dagger$ typically contains strong fluctuation that masks the dominant physics.
Equivalently, the low-energy dynamics can \textit{always} be described by fully quantized effective $\tilde{H}$ upon absorbing rapid fluctuations into dressed particles $\tilde{c}_i^\dagger$.
The desired effective description can be obtained through the bare description, $H^\mathrm{(a)}[\{d_i^\dagger\}]$, of a fictitious `interaction annealed' system with suppressed fluctuation, given $d_i^\dagger$'s resemblance to the dressed particles $\tilde{d}_i^\dagger$ and the ``adiabatic connection'' between $\tilde{d}_i^\dagger$ and $\tilde{c}_i^\dagger$ (see text).}
\vspace{-0.2cm}
\label{schematics}
\end{figure}

\clearpage

\begin{figure}
\centering
\includegraphics[width=0.65\columnwidth]{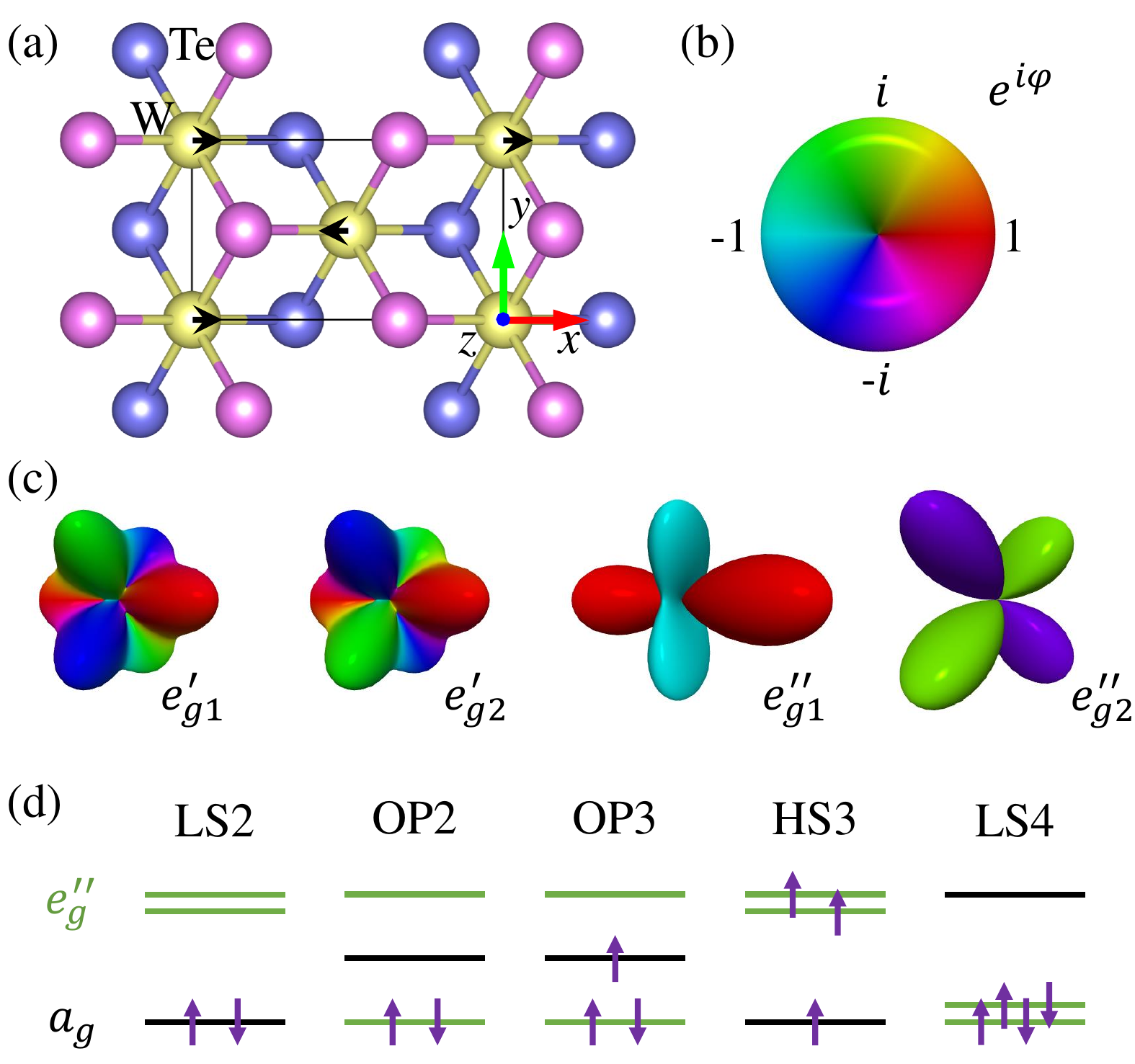}
\vspace{0.5cm}
\caption{Lattice and orbital structures of WTe$_2$. (a) Top view of the $T_d$ lattice structure of WTe$_2$, represented via distortion (black arrows) of the higher symmetry $1T$ lattice structure, 
together with the local coordinate axis of W orbitals.
(b) Color scheme for labeling the phase factor of orbitals.
(c) Symmetry related (degenerate) $e'_g$ orbitals of W~\cite{spherharm_web}, and their superposition, $|e''_{g1}\rangle = \frac{1}{\sqrt{2}}(|e'_{g1}\rangle+|e'_{g2}\rangle)$ and $|e''_{g2}\rangle = \frac{1}{\sqrt{2}}(|e'_{g1}\rangle- |e'_{g2}\rangle)$ emerged in orbital polarized states that spontaneously break the symmetry.
(d) The stable local electronic structures found using LDA+$U$, including low-spin (LS), orbital-polarized (OP), and high-spin (HS) configurations of W ions.}
\vspace{-0.1cm}
\label{WTe2_structure}
\end{figure}

\clearpage

\begin{figure}
\centering
\includegraphics[width=0.9\columnwidth]{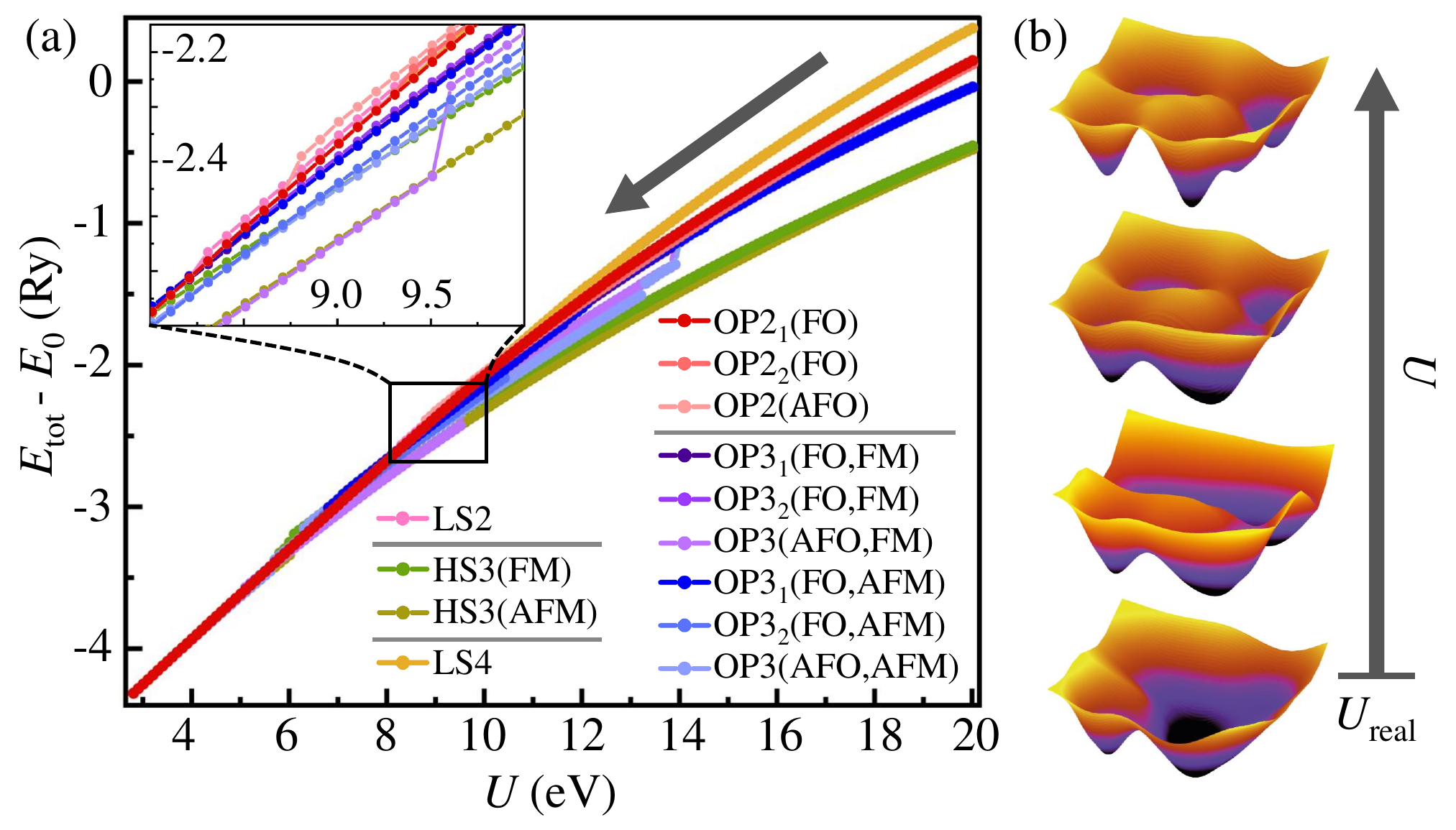}
\vspace{0.5cm}
\caption{Competing electronic structures via interaction annealing. Illustration of competing structures under the symmetric $1T$ structure of WTe$_2$, through a smooth evolution of (a) total energy (for 4 chemical formula units) upon reduction of $U$ and (b) local minimum in the energy contour against density matrices.
The inset highlights the destabilization of some of the structures, corresponding to the vanishing of local energy minimum in (b).
Due to the large energy scale of intra-atomic physics, configurations of distinct local structures separate into four groups, OP2+LS2 (red), HS3 (green), OP3 (blue), and LS4 (yellow) [c.f. Fig.~\ref{WTe2_structure}(d)], but less sensitive to long-range ferro-orbital (FO), anti-ferro-orbital (AFO), ferromagnetic (FM), and antiferromagetic (AFM) long-range orders.}
\vspace{-0.3cm}
\label{fig3}
\end{figure}

\clearpage

\begin{table}[ht!]
\caption{One-body density matrix $\rho_{nn^\prime}$ among the essential W orbitals $n$ in various stable ionic configurations.
(Left column) For a realistic $T_d$ structure with intra-atomic interaction $U=3$ eV, $\rho_{nn^\prime}$ contains fractional occupations (the diagonal elements) in all three orbitals due to strong charge fluctuation.
Upon suppressing fluctuations through increased $U=20$ eV, $\rho_{nn^\prime}$ shows clean occupation of only one of the orbitals, corresponding to an OP2 configuration of Fig.~\ref{WTe2_structure}(d).
(Right columns) For a fictitious system of higher symmetric $1T$ structure and $U=8$ eV, two stable configurations appear similar, but their interaction annealed counterpart reveals qualitatively distinct quantized ionic structures corresponding to an OP2 and a LS4 of Fig.~\ref{WTe2_structure}.}
\begin{tabular}{c|ccc||c|ccc|ccc}
\toprule
$\rho_{nn^\prime}$ & \multicolumn{3}{c||}{$T_d$ OP2} & $\rho_{nn^\prime}$ & \multicolumn{3}{c|}{$1T$ OP2} & \multicolumn{3}{c}{$1T$ LS4} \\ \hline
$U$  & $a_g$ & $e''_{g1}$ & $e''_{g2}$ & $U$ & $a_g$ & $e''_{g1}$ &$e''_{g2}$ &$a_g$ & $e''_{g1}$ &$e''_{g2}$ \\ \hline
& 0.54 &  -0.03 & 0.00 & & 0.45 & -0.01 & 0.00 & 0.46 & -0.01&0.00 \\ 
3 & -0.03 &  0.38 & 0.00 & 8 & -0.01 &  0.58 & 0.00 &-0.01&0.57 &0.00 \\ 
& 0.00 & 0.00 & 0.57 & & 0.00 &  0.00 & 0.53 &0.00 & 0.00 &0.53 \\ \hline
& 0.20 &  0.03 & 0.00 & & 0.12 & -0.08 & 0.00 &0.09 & -0.00 & 0.00 \\
20  & 0.03 & 0.18 & 0.00 & 20 & -0.08 &  0.98 & 0.00 &-0.00 & 0.81 & 0.00 \\
 & 0.00 & 0.00 & 0.96 &  & 0.00 & 0.00 & 0.15 &0.00 & 0.00 & 0.80 \\
\botrule
\end{tabular}
\label{tab1}
\vspace{0.1cm}
\end{table}

\clearpage

\begin{figure}
\centering
\includegraphics[width=0.95\columnwidth]{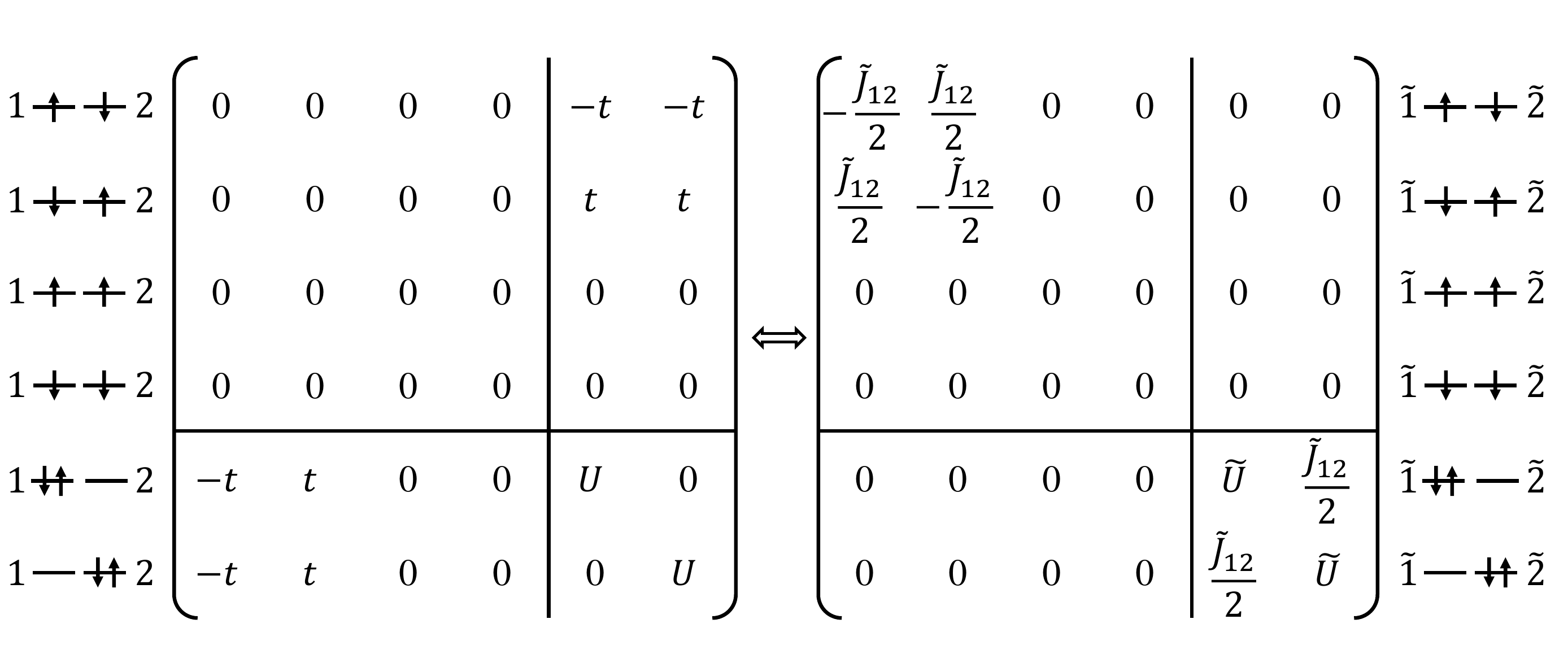}
\caption{
Illustration of the emergent effective description with quantized valence, orbital, and spin structures, using the first-quantized representation of the two-site Hubbard model in its correlated regime ($t\ll U$).
Left panel represents the Hamiltonian in the basis of product states of bare particles $c^\dagger_i$, while the right panel in the dressed particles $\tilde{c}^\dagger_i$.
}
\label{figs1}
\end{figure}

\clearpage

\begin{figure}[!t]
\centering
\includegraphics[width=0.95\textwidth]{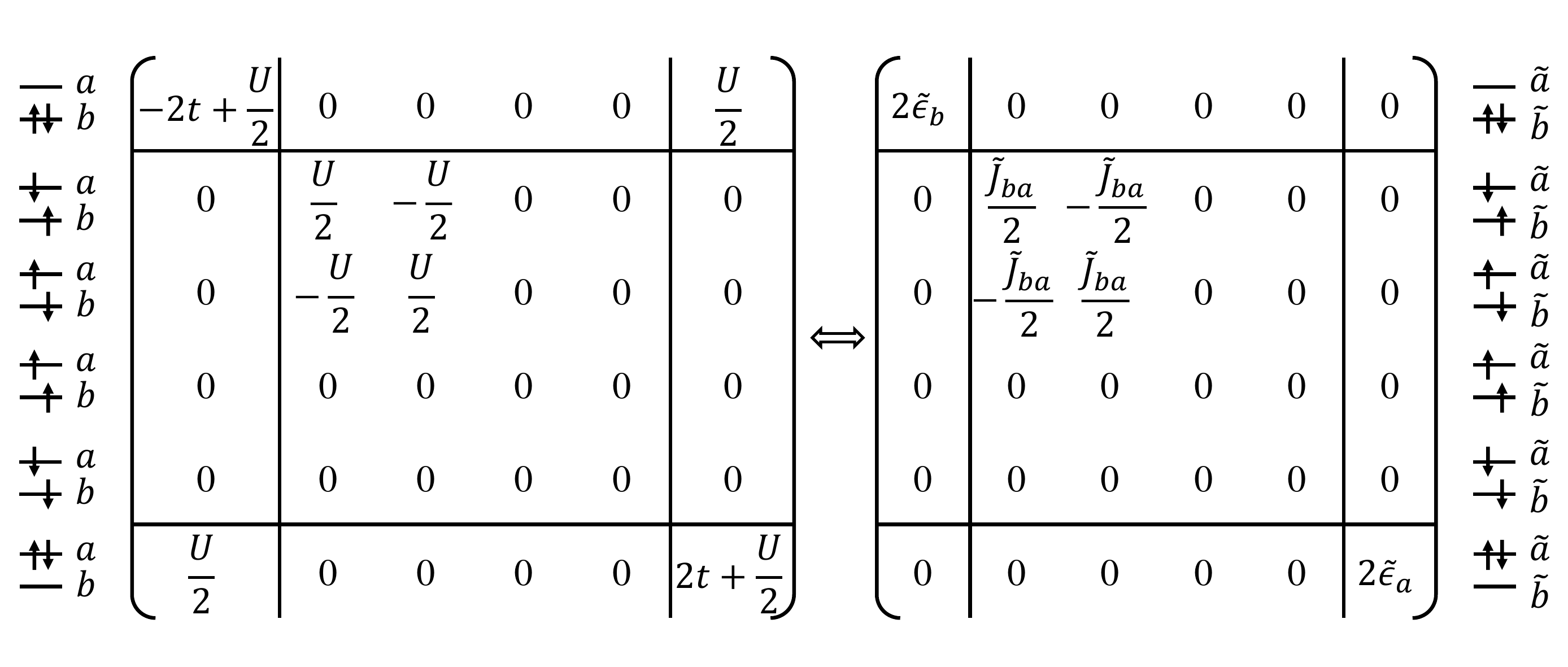}
\caption{
Same as Fig.~\ref{figs1} but for the uncorrelated ($t\gg U$) regime, where the emergent objects resides in the \textit{effective} bonding $\tilde{b}^\dagger_\mu$ and anti-bonding $\tilde{a}^\dagger_\mu$ orbitals.}
\label{Intro_fig2}
\end{figure}

\clearpage

\begin{table}
\centering
\caption{
One-body density matrix $\rho_{nn^\prime}$ among the local W $t_{2g}$ orbitals with index $n$ for different $U \sim 20$ eV configurations in 1$T$ structure.
}
\begin{tabular}[t]{c|ccc|ccc|ccc|ccc}
\toprule
& \multicolumn{3}{c|}{W(1)-$\uparrow$} &\multicolumn{3}{c|}{W(1)-$\downarrow$}& \multicolumn{3}{c|}{W(2)-$\uparrow$} &\multicolumn{3}{c}{W(2)-$\downarrow$}\\  \hline  
$\rho_{nn^\prime}$       & $a_g$  & $e''_{g1}$ &$e''_{g2}$ & $a_g$ & $e''_{g1}$ &$e''_{g2}$  & $a_g$  & $e''_{g1}$ &$e''_{g2}$ & $a_g$ & $e''_{g1}$ &$e''_{g2}$ \\ \hline
                         & 0.98  &-0.00  &-0.00     & 0.98  &-0.00  &-0.00    & 0.98  &-0.00  &-0.00     & 0.98  &-0.00  &-0.00  \\
LS2                      &-0.00  & 0.14  &-0.00     &-0.00  & 0.14  &-0.00    &-0.00  & 0.14  &-0.00     &-0.00  & 0.14  &-0.00  \\
                         &-0.00  &-0.00  & 0.14     &-0.00  &-0.00  & 0.14    &-0.00  &-0.00  & 0.14     &-0.00  &-0.00  & 0.14  \\ \hline
                         & 0.99  & 0.00  & 0.00     & 0.06  & 0.00  & 0.00    & 0.99  & 0.00  & 0.00     & 0.06  & 0.00  & 0.00  \\
HS3(FM)                  & 0.00  & 0.99  &-0.00     & 0.00  & 0.06  & 0.00    & 0.00  & 0.99  &-0.00     & 0.00  & 0.06  & 0.00  \\
                         & 0.00  &-0.00  & 0.99     & 0.00  & 0.00  & 0.06    & 0.00  &-0.00  & 0.99     & 0.00  & 0.00  & 0.06  \\ \hline
                         & 0.99  & 0.00  & 0.00     & 0.06  & 0.00  & 0.00    & 0.06  & 0.00  & 0.00     & 0.99  & 0.00  & 0.00  \\
HS3(AFM)                 & 0.00  & 0.99  & 0.00     & 0.00  & 0.06  & 0.00    & 0.00  & 0.06  & 0.00     & 0.00  & 0.99  & 0.00  \\
                         & 0.00  & 0.00  & 0.98     & 0.00  & 0.00  & 0.06    & 0.00  & 0.00  & 0.06     & 0.00  & 0.00  & 0.98  \\ \hline
                         & 0.09  &-0.00  &-0.00     & 0.09  &-0.00  &-0.00    & 0.09  &-0.00  &-0.00     & 0.09  &-0.00  &-0.00  \\
LS4                      &-0.00  & 0.81  & 0.00     &-0.00  & 0.81  & 0.00    &-0.00  & 0.81  & 0.00     &-0.00  & 0.81  & 0.00  \\
                         &-0.00  & 0.00  & 0.80     &-0.00  & 0.00  & 0.80    &-0.00  & 0.00  & 0.80     &-0.00  & 0.00  & 0.80  \\ \hline
                         & 0.12  &-0.08  & 0.00     & 0.12  &-0.08  & 0.00    & 0.12  &-0.08  & 0.00     & 0.12  &-0.08  & 0.00  \\
OP2$_1$(FO)              &-0.08  & 0.98  & 0.00     &-0.08  & 0.98  & 0.00    &-0.08  & 0.98  & 0.00     &-0.08  & 0.98  & 0.00  \\
                         & 0.00  & 0.00  & 0.15     & 0.00  & 0.00  & 0.15    & 0.00  & 0.00  & 0.15     & 0.00  & 0.00  & 0.15  \\ \hline
                         & 0.12  & 0.04  & 0.00     & 0.12  & 0.04  & 0.00    & 0.12  & 0.04  & 0.00     & 0.12  & 0.04  & 0.00  \\
OP2$_2$(FO)              & 0.04  & 0.18  & 0.00     & 0.04  & 0.18  & 0.00    & 0.04  & 0.18  & 0.00     & 0.04  & 0.18  & 0.00  \\
                         & 0.00  & 0.00  & 0.98     & 0.00  & 0.00  & 0.98    & 0.00  & 0.00  & 0.98     & 0.00  & 0.00  & 0.98  \\ \hline
                         & 0.11  & 0.09  & 0.00     & 0.11  & 0.09  & 0.00    & 0.15  & 0.01  & 0.00     & 0.15  & 0.01  & 0.00  \\
OP2(AFO)                 & 0.09  & 0.97  & 0.00     & 0.09  & 0.97  & 0.00    & 0.01  & 0.20  & 0.00     & 0.01  & 0.20  & 0.00  \\
                         & 0.00  & 0.00  & 0.16     & 0.00  & 0.00  & 0.16    & 0.00  & 0.00  & 0.99     & 0.00  & 0.00  & 0.99  \\ \hline   
                         & 0.97  &-0.03  & 0.00     & 0.07  &-0.02  & 0.00    & 0.97  &-0.03  & 0.00     & 0.07  &-0.02  & 0.00  \\
OP3$_1$(FO,FM)           &-0.03  & 0.94  & 0.00     &-0.02  & 0.93  & 0.00    &-0.03  & 0.94  & 0.00     &-0.02  & 0.93  & 0.00  \\
                         & 0.00  & 0.00  & 0.09     & 0.00  & 0.00  & 0.08    & 0.00  & 0.00  & 0.09     & 0.00  & 0.00  & 0.08  \\ \hline
                         & 0.97  & 0.07  & 0.00     & 0.07  & 0.00  & 0.00    & 0.97  & 0.07  & 0.00     & 0.07  & 0.00  & 0.00  \\
OP3$_2$(FO,FM)           & 0.07  & 0.09  & 0.00     & 0.00  & 0.08  & 0.00    & 0.07  & 0.09  & 0.00     & 0.00  & 0.08  & 0.00  \\
                         & 0.00  & 0.00  & 0.93     & 0.00  & 0.00  & 0.93    & 0.00  & 0.00  & 0.93     & 0.00  & 0.00  & 0.93  \\ \hline
                         & 0.98  &-0.02  & 0.00     & 0.07  &-0.00  & 0.00    & 0.98  &-0.06  & 0.00     & 0.07  &-0.00  & 0.00  \\
OP3(AFO,FM)              &-0.02  & 0.95  & 0.00     &-0.00  & 0.93  & 0.00    &-0.06  & 0.10  & 0.00     &-0.00  & 0.08  & 0.00  \\
                         & 0.00  & 0.00  & 0.09     & 0.00  & 0.00  & 0.08    & 0.00  & 0.00  & 0.96     & 0.00  & 0.00  & 0.94  \\ \hline
                         & 0.97  &-0.03  & 0.00     & 0.07  &-0.00  & 0.00    & 0.07  &-0.00  & 0.00     & 0.97  &-0.03  & 0.00  \\
OP3$_1$(FO,AFM)          &-0.03  & 0.95  & 0.00     &-0.00  & 0.93  & 0.00    &-0.00  & 0.93  & 0.00     &-0.03  & 0.95  & 0.00  \\
                         & 0.00  & 0.00  & 0.09     & 0.00  & 0.00  & 0.08    & 0.00  & 0.00  & 0.08     & 0.00  & 0.00  & 0.09  \\ \hline
                         & 0.99  & 0.03  & 0.00     & 0.07  & 0.00  & 0.00    & 0.07  & 0.00  & 0.00     & 0.99  & 0.03  & 0.00  \\
OP3$_2$(FO,AFM)          & 0.03  & 0.09  & 0.00     & 0.00  & 0.08  & 0.00    & 0.00  & 0.08  & 0.00     & 0.03  & 0.09  & 0.00  \\
                         & 0.00  & 0.00  & 0.94     & 0.00  & 0.00  & 0.91    & 0.00  & 0.00  & 0.91     & 0.00  & 0.00  & 0.94  \\ \hline
                         & 0.98  &-0.03  & 0.00     & 0.07  & 0.00  & 0.00    & 0.07  &-0.00  &-0.00     & 0.99  & 0.00  & 0.00  \\
OP3(AFO,AFM)             &-0.03  & 0.94  & 0.00     & 0.00  & 0.93  & 0.00    &-0.00  & 0.08  & 0.00     & 0.00  & 0.09  & 0.00  \\
                         & 0.00  & 0.00  & 0.09     & 0.00  & 0.00  & 0.08    &-0.00  & 0.00  & 0.93     & 0.00  & 0.00  & 0.94  \\                  
\botrule
\end{tabular}
\label{tabs1}
\end{table}

\clearpage

\begin{figure}
\centering
\includegraphics[width=1\columnwidth]{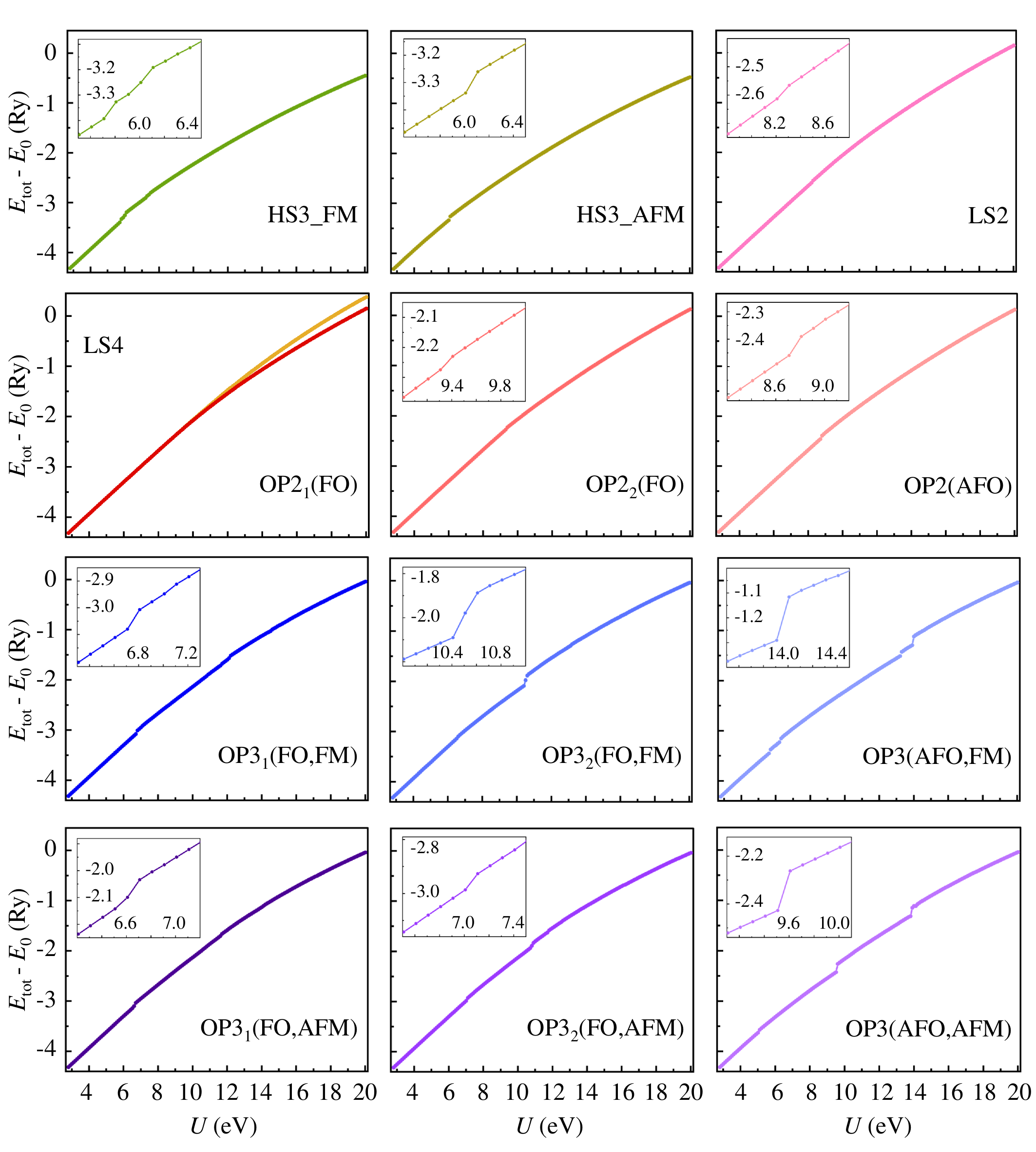}
\caption{(color online) 
The smooth evolution of total energy upon reduction of the interaction strength $U$ in different configurations under $1T$ structure of WTe$_2$.
As the value of $U$ gradually decreases, the total energy evolves smoothly until some configurations become unstable and fall to other more stable ones, as indicated by the abrupt `jump' in the total energy curves.
}
\label{detail_annealing}
\end{figure}

\clearpage

\begin{figure}
\centering
\includegraphics[width=0.55\columnwidth]{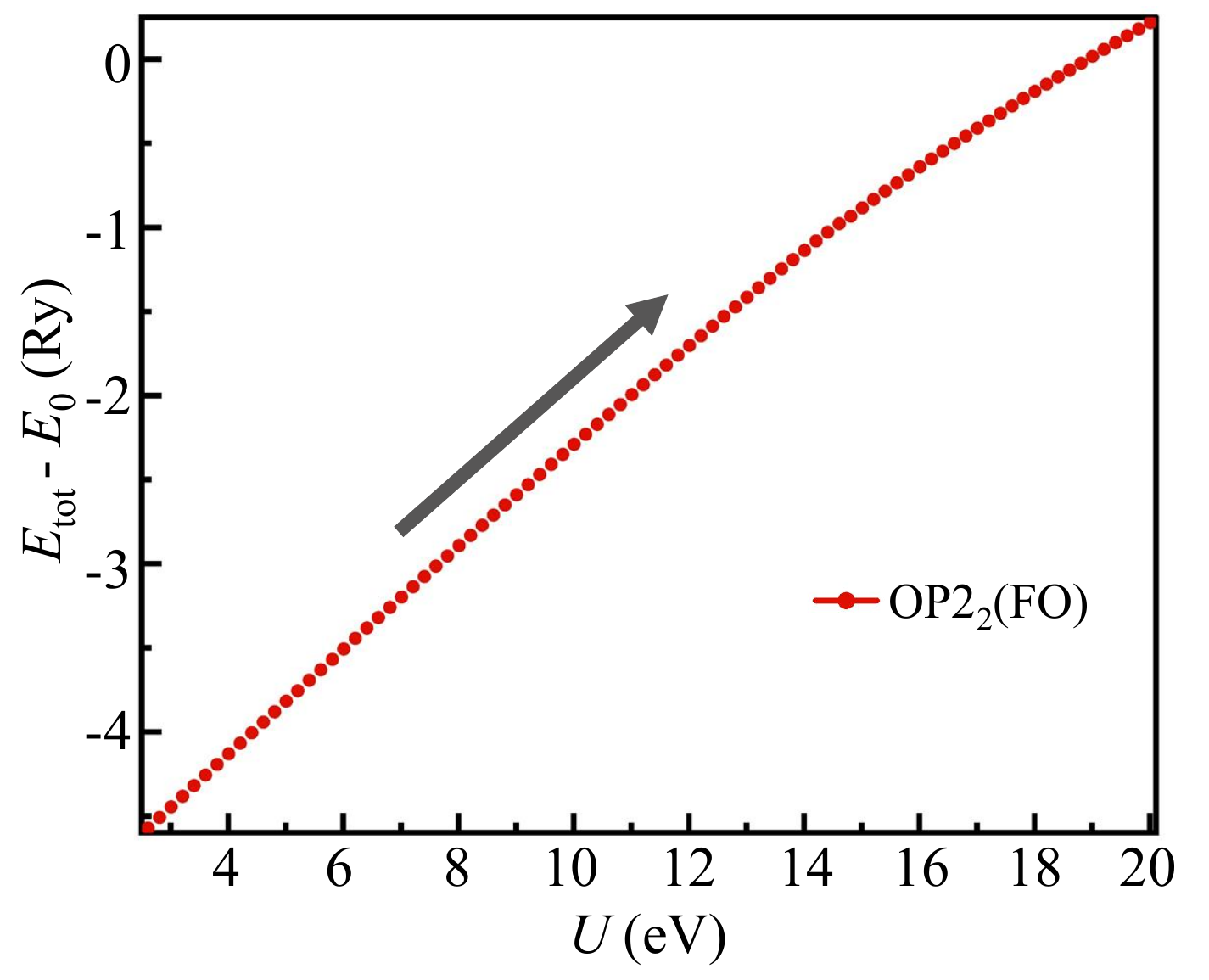}
\caption{(color online) Stable configuration of WTe$_2$ and its adiabatic connection upon `heating up' the interaction strength $U$ in $T_d$ structure.
As $U$ slowly increases from the realistic value $\sim3$ eV to large limit $\sim20$ eV, the configuration is clearly the OP2$_2$ with FO ordering, corresponding to two electrons both occupying one of the two degenerate orbitals $e''_{g2}$ with long-range ferro-orbital order.
}
\label{Td_annealing}
\end{figure}

\clearpage

\begin{table}
\caption{
Representative off-diagonal matrix elements, $\rho_{pd}$, of one-body density matrix between Te $p$-orbitals and W $d$-orbitals in the $T_d$ structure, in comparison with the corresponding diagonal elements $\rho_{pp}$ and $\rho_{dd}$.}
\begin{tabular}{cc|ccc|ccc}
\toprule
&&&$U \sim 3$ &&&$U \sim 20$ &\\ \hline
Te-$p$ & W-$d$ & $\rho_{pd}$ & $\rho_{pp}$ &$\rho_{dd}$ &$\rho_{pd}$ & $\rho_{pp}$ &$\rho_{dd}$ \\ \hline
$p_x$ & $d_{a_g}$ & 0.20 & 0.69 & 0.54 & 0.17 & 0.79 & 0.20 \\
$p_x$ & $d_{e_g}$ & 0.18 & 0.67 & 0.41 & 0.16 & 0.70 & 0.28 \\
$p_y$ & $d_{a_g}$ & 0.14 & 0.63 & 0.54 & 0.12 & 0.69 & 0.20 \\
$p_y$ & $d_{e_g^{\prime\prime}}$ & 0.14 & 0.64 & 0.38 & 0.12 & 0.70 & 0.19\\
$p_z$ & $d_{e_g}$ & 0.11 &  0.68 & 0.41  & 0.11 & 0.67 & 0.28  \\
\botrule
\end{tabular} 
\label{tabs2}
\vspace{-8pt}
\end{table}

\clearpage

\begin{table}[!h]
\caption{\label{tab:struct} Estimated octahedral distortion from ionic radius~\cite{shannon1976revised} for $d^2$ and $d^3$ configurations in WTe$_2$ with $T_{d}$ structure, in comparison with the experimental parameter.
Distortion is evaluated by the shortest W-Te bond length, $\Delta=\sum_{i=1}^6\left|d_i-d_{\text{mean}}\right|$, and distortion parameters, $\Sigma=\sum_{i=1}^{12}\left|90-\phi_i\right|$, where $\phi_i$ denotes the intersection angle $B_{i_1}-A-B_{i_2}$ of octahedron $AB_6$.}
\label{tab:distortion_struct}
\vspace{5pt} 
\begin{tabular}{cccc}
\toprule
\vspace{5pt} 
& \begin{tabular}[c]{@{}c@{}}$T_d$ structure\\ experiment~\cite{Tao_PhysRevB2020}\end{tabular} & \begin{tabular}[c]{@{}c@{}}$d^2$\\ configuration\end{tabular} & \begin{tabular}[c]{@{}c@{}}$d^3$\\ configuration\end{tabular} \\
\hline
$\Delta (\mathbf{\AA})$ & 0.306 & 0.319 & 0.073 \\
$\Sigma (^{\circ})$ & 145.200 & 162.963 & 204.611\\
\botrule
\end{tabular}
\vspace{-0.6cm}
\end{table}

\end{document}